\documentclass[useAMS,usenatbib]{mnras}

\usepackage{graphicx}
\usepackage{epstopdf}
\usepackage{subcaption}
\usepackage{caption}
\usepackage{amsmath}
\usepackage{gensymb}
\usepackage[normal]{threeparttable}
\usepackage{calc}
\usepackage{pdflscape}
\usepackage{rotating}
\usepackage{afterpage}

\title[]{An ASKAP survey for H{\sc i} absorption towards dust-obscured quasars}
\author[]{M.~Glowacki$^{1,2,3,4,5}$\thanks{E-mail:
marcin@idia.ac.za}, J.R.~Allison$^{6,7}$, V.A.~Moss$^{8,1}$, E.K.~Mahony$^{3}$, E.M.~Sadler$^{1,7}$,\newauthor J.R.~Callingham$^{8}$, S.L.~Ellison$^{9}$, M.T.~Whiting$^{3}$, J.D. Bunton$^{3}$, A.P.~Chippendale$^{3}$, \newauthor I.~Heywood$^{6,10}$, D.~McConnell$^{3}$, W.~Raja$^{3}$, M.A.~Voronkov$^{3}$\\
$^{1}$Sydney Institute for Astronomy, School of Physics A28, University of Sydney, NSW 2006, Australia\\
$^{2}$ARC Centre of Excellence for All-sky Astrophysics (CAASTRO), Australia\\
$^{3}$CSIRO Astronomy \& Space Science, PO Box 76, Epping NSW 1710, Australia\\
$^{4}$Inter-University Institute for Data Intensive Astronomy, University of Cape Town, South Africa\\
$^{5}$Department of Physics and Astronomy, University of the Western Cape, Robert Sobukwe Road, Bellville 7535, South Africa\\
$^{6}$Sub-Dept. of Astrophysics, Department of Physics, University of Oxford, Denys Wilkinson Building, Keble Rd., Oxford, OX1 3RH, UK\\
$^{7}$ARC Centre of Excellence for All Sky Astrophysics in 3 Dimensions (ASTRO 3D)\\
$^{8}$ASTRON, the Netherlands Institute for Radio Astronomy, Postbus 2, 7990 AA, Dwingeloo, The Netherlands\\
$^{9}$Department of Physics \& Astronomy, University of Victoria, Finnerty Road, Victoria, British Columbia, V8P 1A1, Canada\\
$^{10}$Department of Physics \& Electronics, Rhodes University, PO Box 94, Grahamstown, 6140, South Africa}

\begin{document}


\pagerange{\pageref{firstpage}--\pageref{lastpage}} \pubyear{2002}

\maketitle

\label{firstpage}

\begin{abstract} Obscuration of quasars by accreted gas and dust, or dusty intervening galaxies, can cause active galactic nuclei (AGN) to be missed in optically-selected surveys. Radio observations can overcome this dust bias. In particular, radio surveys searching for H{\sc i} absorption inform us on how the AGN can impact on the cold neutral gas medium within the host galaxy, or the population of intervening galaxies through the observed line of sight gas kinematics. We present the results of a H{\sc i} absorption line survey at $0.4 < z < 1$ towards 34 obscured quasars with the Australian SKA Pathfinder (ASKAP) commissioning array. We detect three H{\sc i} absorption lines, with one of these systems previously unknown. Through optical follow-up for two sources, we find that in all detections the H{\sc i} gas is associated with the AGN, and hence that these AGN are obscured by material within their host galaxies. Most of our sample are compact, and in addition, are either gigahertz peaked spectrum (GPS), or steep spectrum (CSS) sources, both thought to represent young or recently re-triggered radio AGN. The radio spectral energy distribution classifications for our sample agree with galaxy evolution models in which the obscured AGN has only recently become active. Our associated H{\sc i} detection rate for GPS and compact SS sources matches those of other surveys towards such sources. We also find shallow and asymmetric H{\sc i} absorption features, which agrees with previous findings that the cold neutral medium in compact radio galaxies is typically kinematically disturbed by the AGN. \end{abstract}


\section{Introduction}

\subsection{Quasar obscuration}
Obscuration of quasars occurs in two distinct ways. One is due to gas and dust within an intervening galaxy along the line of sight, which affects our ability to detect the background quasar through optical surveys. Dusty intervening galaxies obscuring quasars have been detected in, for example, \cite{Noterdaeme2010} and \cite{Srianand2008}. However, dusty absorbers do not appear to dominate the statistics of intervening absorbers \citep{Ellison2001, Ellison2004}. Gravitationally lensed quasars can be also reddened by dust within the lensing galaxy \citep{Malhotra1997,Gregg2002}. 

The other cause of quasar obscuration is due to gas and dust within the host galaxy itself. In this model of galaxy evolution \cite[][and references within]{Hopkins2006,Glikman2007,Hopkins2008}, a galaxy merger event is the primary method of contributing the dusty material which obscures the active galactic nucleus (AGN), as seen in \cite{Satyapal2014,Ellison2015,Weston2017}.  Cold-mode accretion of the obscuring material onto the AGN continues until AGN feedback is triggered, which dispels the obscuring material \citep{Koss2010,Ellison2011}. The optically obscured period is relatively short-lived \cite[timescale of a few million years or $\sim$15--20\% of the unobscured quasar phase;][]{Glikman2012}, and hence galaxies in this evolutionary phase are comparatively rare.

The short time scale of the obscured evolution phase hampers our ability to study the AGN impact on the cold neutral medium. Difficulty with obtaining sufficient optical spectral information for obscured quasars, resulting in such objects falling out of optically selected samples, also restricts our ability to understand the evolution of the AGN population across redshift space \cite[see review by][]{Hickox2018}. This missing information can create a mismatch with the expected population of obscured galaxies in galaxy evolution models, such as the model of \cite{Hopkins2006}. Another motivation to study AGN obscured by associated gas and dust is to understand the impact of the AGN on this obscuring material \cite[e.g. in extreme cases, acceleration of gas by radio jets associated with the AGN;][]{Morganti2013}. The available neutral gas content in recent merger systems are typically a factor of three higher than other galaxies with the same stellar mass \citep{Dutta2018,Ellison2018}. This is hence a crucial phase of galaxy evolution.

\subsection{Finding obscured quasars}
\subsubsection{Infrared wavelengths}
Other wavelengths than the optical are employed to study dust-obscured quasars. Infrared selection methods from e.g. the Widefield Infrared Survey Explorer mission \citep[WISE;][]{Wright2010} can trace the warm dust in the obscuring torus of AGN, and can also identify optically faint quasars \cite[e.g.][]{Richards2015}. AGN are often identified with redder $W1-W2$ (3.4~--~4.6~$\mu$m) colour \citep{Stern2012,Assef2013,Blecha2018}. 

Meanwhile, the colour of objects with redder mid-infrared $W2-W3$ (4.6~--~12~$\mu$m) magnitudes are attributed to warm continuum emission from polycyclic aromatic hydrocarbons (PAHs), which is traced by the W3 band. PAHs are found in regions of recent star formation \citep{Lee2013,Cluver2014}, and such star forming regions can be detected through spectroscopic studies. AGN with red WISE colours may be associated with galaxy mergers, which can be the source of high obscuration \citep{Blecha2018}. 
\subsubsection{Radio wavelengths}
Radio-selected samples of quasars avoid the limitations introduced by dust for optical surveys \citep{Ellison2001}, as well as the issue of confusion of obscured quasars with low-mass stars in optical colour selection. \cite{Webster1995} suggested that the wide range of optical colours for radio-selected quasars is due to dust along the line of sight, and estimated that up to 80\% of quasars could have been missed by optical surveys. \cite{White2003} found that the population of radio-selected quasars is dominated by previously undetected red, heavily obscured objects. These claims have since been found to be an overestimate; \cite{Gregg2001} estimated that between 10\% and 20\% of the radio-loud quasar population is reddened by dust in the host galaxy. It was found by \cite{Pontzen2009} that 7\% of quasars with damped Lyman~$\alpha$ absorbers (DLA) are expected to be missing from optically selected samples. \cite{Francis2000} showed that only about 10\% of the radio-selected quasars showed signs of obscuration, while \cite{Whiting2001} demonstrated that the majority were reddened by non-thermal emission from the radio jet. Nonetheless, if DLA absorbers associated with merger systems are systematically obscured, then this  population could be missed without radio selection methods. 

The spectral energy distribution (SED) at radio wavelengths also provides information on the morphology and age of radio AGN. Radio sources with a peak in their SED in the MHz or GHz regime (e.g. gigahertz peaked spectrum sources; GPS) or which increase in flux density with decreasing frequency and are compact (compact steep spectrum sources; CSS) are believed to be young or recently re-triggered radio AGN \cite[e.g.][]{ODea1991}. GPS radio galaxies are compact ($\leq$~1~kpc in size), more so than CSS sources ($\sim$1--10 kpc). 

GPS or CSS radio galaxies that are obscured would agree with galaxy evolution models \cite[e.g.][]{Hopkins2006}, in which obscured quasars represent a short transitional period in which the AGN is `switching on' following a fuelling stage by obscuring material \citep{Blecha2018}. An alternative model is that these sources are `frustrated' - rather than being young, the radio source is confined to small spatial scales by a high nuclear plasma density. \cite{Lonsdale2016} found that the majority of a sample of 155 optically obscured quasars were compact, steep-spectrum, and sub-galactic in scale, and their SEDs consistent with young radio source ages.

\subsection{H{\sc i} absorption}
The 21~cm hyperfine transition of atomic hydrogen (H{\sc i}) traces the kinematics of the neutral gas reservoir within galaxies. Surveys of H{\sc i} in emission \cite[e.g.][]{Meyer2004,Giovanelli2005,Verheijen2007,Catinella2008,Freudling2011} are limited beyond redshifts $z$~$\sim$~0.3 due to the faintness of the line. The record is currently $z$~=~0.376 \citep{Fernandez2016}. One method to detect H{\sc i} at higher redshifts is the 21~cm transition in absorption toward background radio AGN \cite[e.g.][]{Carilli1998,Chengalur2000,Kanekar2003,Vermeulen2003,Curran2006,Gupta2006,Srianand2008,Allison2012,Gereb2015,Glowacki2017,Maccagni2017}. 

In 21~cm absorption, we can search along the line of sight, and from the H{\sc i} kinematic information, investigate the impact of the AGN on the cold neutral gas during this obscured period. Intervening 21~cm H{\sc i} absorption, meanwhile, can inform us about the population of DLA systems along the line of sight to radio-loud AGN. From these galaxies we can learn more on the distribution, quantity and evolution of cold neutral gas in galaxies with redshift from 21~cm studies, such as on the potential evolution of the gas spin temperature with redshift \citep{Ellison2012,Kanekar2014,Curran2018}. 

Absorption line studies have been shown to uncover cold gas in dust-obscured galaxies. This has been demonstrated through H$_{2}$ and C{\sc i} absorption detections in DLA systems \citep{Ge1997,Ge1999}. This work supports previous findings that DLAs can be the source of reddening of background quasars \citep{Kulkarni1997}. In another example, \cite{Ledoux2015} searched for C\,I in quasar spectra, and state their sample of C\,I absorbers provides ideal targets for studying cold gas towards high redshift quasars.

Similarly, 21~cm H{\sc i} absorption studies provide us with a method to study obscured quasars. However, while associated and intervening H{\sc i} absorption have been detected towards obscured quasars in previous studies, these are limited in sample size. \cite{Carilli1998} found an 80\% detection rate in H{\sc i} absorption towards a sample of 5 reddened quasars that were bright in the radio \cite[see also][]{Carilli1999}. This was a deep survey with integrated optical depths of $\sim$1\% achieved. Another associated H{\sc i} absorption detection was made by \cite{Ishwara-Chandra2003} towards 3C190, classified as a red quasar due to a steep fall-off of the optical to infrared continuum. Detections of intervening H{\sc i} absorption were made by \cite{Srianand2008} toward two red quasars, which exhibited strong, intervening Mg\,{\sc II} absorption lines.

There are multiple factors that can decrease the chance of detecting H{\sc i} absorption, such as a lower covering factor of the radio source when it is spatially extended, or a low radio-flux density or short on-source integration time. One primary issue for radio surveys, radio frequency interference (RFI), has restricted these and particularly H{\sc i} absorption studies towards obscured AGN at higher redshift. For instance, one of the H{\sc i} absorption features detected in \cite{Carilli1998} was partially corrupted by RFI. 

The Australian Square Kilometre Array Pathfinder \cite[ASKAP;][]{Deboer2009, Johnston2009, Schinckel2012} will address these issues, in part due to the radio-quiet nature of the site. The wide 300~MHz instantaneous bandwidth of the telescope allows for a large redshift space to be probed, and the large field-of-view (30 square degrees) increases survey speed. As a result, the First Large Absorption Survey in H{\sc i} (FLASH; Sadler et al., in prep) with ASKAP will target more than 100,000 sightlines to bright radio sources in the entire southern sky in the redshift range 0.4~$<$~$z$~$<$~1 for both intervening and associated H{\sc i} absorption. FLASH is estimated to improve on other H{\sc i} absorption surveys by over two orders of magnitude in the number of sightlines searched for H{\sc i} gas. The other SKA Pathfinders, the South African MeerKAT telescope \citep{Booth2012} and the Westerbork APERture Tile in Focus \cite[APERTIF;][]{Oosterloo2009}, will complement FLASH through the MeerKAT Absorption Line Survey \cite[MALS;][]{Gupta2017} and the Search for H{\sc i} absorption with APERTIF (SHARP\footnote{\url{http://www.astron.nl/astronomy-group/apertif/science-projects/sharp-search-hi-absorption-apertif/sharp}}) respectively at different redshift ranges \cite[see also][]{Maccagni2017}.

We present the findings of a H{\sc i} absorption survey towards obscured radio quasars during the commissioning stage of ASKAP with the Boolardy Engineering Test Array \cite[BETA;][]{Hotan2014,McConnell2016}. The radio sample was selected either due to known optical faintness from SuperCOSMOS \citep{Hambly2001}, or via additional mid-infrared \cite[WISE;][]{Wright2010} colour information (see Section~\ref{sec:2}). We consider the spectral energy distribution properties of our sources to investigate whether these obscured quasars are compact and hence potentially young or re-triggered in their radio activity. To verify whether we could detect H{\sc i} absorption towards optically faint sources in our sample with no redshift information, we used a photometric redshift indicator method \citep{Glowacki2017b} to estimate their redshifts. In Section~\ref{sec:3} we summarise the observations made. In Section~\ref{sec:4} we report the results of the H{\sc i} absorption survey and the individual detections. In Section~\ref{sec:5} we discuss and analyse our H{\sc i} absorption detection rates, and further analyse the SED properties of our sample. 



\section{Sample properties}\label{sec:2}

\subsection{Sample selection}

\begin{table*}
\normalsize
 \centering
  \caption{Summary of the sample selection. Bright radio candidate quasars south of declination +5$\degree$ and without known redshifts $<$~0.4 are selected from the sample of 518 extragalactic radio sources of \citet{Kuhr1981}. These 34 sources make up two sub-samples; one of optically faint sources through SuperCOSMOS information with no further redshift restriction, and the other mid-infrared reddened sources through WISE colour information, with additional limits on the redshift. An overlap of 7 sources exists between these two sub-samples. The last column gives the number of sources we were able to search for associated H{\sc i} absorption, with the bracketed number sources with spectroscopic redshifts.}
  \begin{tabular}{lclc}
  \hline
Sample summary & No. of sources & Selection criteria & Associated abs.\\
 \hline
 \citealt{Kuhr1981} & 518 & -- & -- \\
 \hline
-- Radio-bright QSO sample & 34 & $\geq$1~Jy in NVSS or SUMSS & 24 (12) \\
    (Table~\ref{tab:sourcelist}) & & South of declination +5$\degree$ \\
  & & No known redshift, or $z$~$>$~0.4\\
    & & Satisfies criteria of at least one sub-sample below\\
   \hline
  --- Optically faint sub-sample & 28 (of 34) & R-band magnitude $\geq$~20~mag & 21 (9) \\
  \hline 
--- WISE selected sub-sample & 13 (of 34) & WISE colour W2--W3~$>$~3.5 & 10 (5)\\
  & & Radio AGN redshift OR intervening non-H{\sc i}  \\
  & & absorption line redshift within range 0.4~$<$~$z$~$<$~1\\
\hline
\end{tabular}
\label{tab:samplesum}
\end{table*}

\begin{table*}
{\centering
\caption{Summary of radio sources selected for observation with ASKAP-BETA. Here we list the full sample as three groups which (a) are searchable for associated H{\sc i} absorption in our survey, (b) have no known redshift or only photometric redshift estimations in the literature, and (c) can only be searched for intervening H{\sc i} absorption. The 1.4 and 0.8~GHz flux densities, S$_{1.4}$ and S$_{0.8}$, come from NVSS and SUMSS respectively \citep{Condon1998,Mauch2003}. We selected sources with flux densities greater than 1~Jy at these frequencies. In the Sample column, Opt represents the optically faint sub-sample, WISE the WISE selected sources, and O+W for objects in both sub-samples, which are defined in Sections~\ref{sec:optsample} and \ref{sec:wisesample}. In the RSC (radio spectral classification) column we classify objects as peaked spectrum (PS), steep spectrum (SS) and flat spectrum (FS) sources (Section~\ref{sec:sed}).}
\label{tab:sourcelist}
\setlength{\tabcolsep}{5pt}
\begin{tabular}{lrrrllccrrrrcr}
\hline
Source                     & RA          & Dec         & S$_{1.4}$ & S$_{0.8}$ & z     & ref        & Sample & W1     & W2     & W3                             & R        & ref  & RSC    \\
                           & J2000       & J2000       & Jy        & Jy         &       &            &        & mag    & mag    & mag                            & mag      &      &      \\
\hline
\multicolumn{6}{l}{(a) Objects with spectroscopic redshift $0.4<z<1$ }\\
\hline
PKS\,0022$-$423                & 00:24:42.95 & -42:02:03.5 & -         & 1.77      & 0.937 & b          & Opt     & 14.6 & 13.4  & 10.0                          & 20.1   & c & PS\\
3C\,038                      & 01:20:27.12 & -15:20:16.9 & 5.08      & -         & 0.565 & e          & WISE     & 14.9 & 14.5 & 10.7                         & 19.0   & c  & SS \\
PKS\,0235$-$19                 & 02:37:43.78 & -19:32:35.4 & 4.62      & -         & 0.62  & e          & WISE     & 14.2 & 12.9 & 9.3                          & 19.6   & c & SS  \\
PKS\,0252$-$71                 & 02:52:46.26 & -71:04:35.9 & -         & 8.55      & 0.568 & f          & Opt     & 15.1 & 14.7  & 11.5                         & 20.6     & a  & PS \\
PKS\,0408$-$65                 & 04:08:20.28 & -65:45:08.5 & -         & 24.44     & 0.962 & h          & O+W  & 15.7 & 14.9 & 10.3                         & 21.4     & a & PS  \\
PKS\,0420$-$014                & 04:23:15.80 & -01:20:33.1 & 2.73      & -         & 0.916 & g$\dagger\ddagger$  & WISE     & 11.5 & 10.4 & 6.7                           & 19.8   & c  & FS \\
PKS\,0500+019               & 05:03:21.20 & +02:03:04.7 & 2.25      & -         & 0.585 & j$\ddagger$   & Opt     & 14.5 & 13.5 & 10.6                         & 20.8   & c  & PS \\
PKS\,1622$-$253                & 16:25:46.89 & -25:27:38.3 & 2.52      & -         & 0.786 & l          & Opt     & 12.5 & 11.4 & 8.5                          & 20.8   & c  & SS \\
PKS\,1740$-$517                & 17:44:25.45 & -51:44:43.8 & -         & 8.15      & 0.44  & -          & Opt     & 14.4 & 13.3 & 10.5                         & 20.8     & a & PS  \\
PKS\,2008$-$068                & 20:11:14.23 & -06:44:03.4 & 2.60      & -         & 0.547 & u          & Opt     & 15.0 & 14.6 & 12.5                         & 21.2    & q  & PS \\
PKS\,2331$-$41                 & 23:34:26.13 & -41:25:25.8 & -         & 8.73      & 0.907 & l          & O+W  & 14.8 & 14.1 & 10.3                         & 20.6   & c & SS  \\
\hline 
\multicolumn{8}{l}{(b) Objects without a reliable spectroscopic redshift (some have photometric redshifts)}\\
\hline
PKS\,0042$-$35                 & 00:44:41.47 & -35:30:32.8 & 2.56         & 3.81      & 0.98*  & a         & O+W  & 16.8 & 15.6 & 10.7                         & 20.1   & c  & SS \\
3C\,118                      & 04:31:07.09 & +01:12:55.5 & 1.65      & -         & -     & -          & Opt     & 15.0 & 14.5 & 10.8                         & 20.1   & c & SS  \\
PKS\,1213$-$17                 & 12:15:46.75 & -17:31:45.4 & 1.84      & -         & -     & -          & Opt     & -      & -      & -                              & -        & -  & SS \\
3C\,275                      & 12:42:19.67 & -04:46:20.7 & -         & 8.92      & 0.56*  & a         & O+W  & 15.0 & 13.4 & 9.4                           & 20.3   & a & SS  \\
PKS\,1829$-$718                & 18:35:37.20 & -71:49:58.2 & -         & 3.99      & -     & -          & Opt     & 15.8 & 15.4 & $>$12.7  & $\sim$23 & p  & PS  \\
PKS\,1936$-$623                & 19:41:21.76 & -62:11:21.0 & -         & 1.80      & -     & -          & Opt     & 13.9 & 12.5  & 9.7                          & 20.4   & c  & SS \\
PKS\,2032$-$35                 & 20:35:47.61 & -34:54:01.1 & -         & 8.92      & 0.56*  & a         & O+W  & 14.5 & 13.2 & 9.6                          & 20.4   & c  & SS \\
PKS\,2140$-$81                 & 21:47:23.62 & -81:32:08.6 & -         & 5.09      & 0.64*  & a         & Opt     & 14.5 & 13.6 & 10.8                         & 20.7   & c  & SS \\
PKS\,2150$-$52                 & 21:54:07.28 & -51:50:18.1 & -         & 5.26      & 0.79*  & a         & O+W  & 15.5 & 14.5 & 10.2                         & 21.4     & a & SS  \\
PKS\,2307$-$282                & 23:10:07.40 & -27:57:51.6 & 1.02      & -         & -     & -          & Opt     & 15.3 & 14.8 & 12.5                         & $>$20.0        & -  & SS \\
PKS\,2323$-$40                 & 23:26:34.30 & -40:27:15.5 & -         & 5.24      & 0.81*  & a         & O+W  & 16.6 & 15.0 & 10.6                         & 21.5     & a  & SS \\
PKS\,2333$-$528                & 23:36:12.14 & -52:36:21.9 & -         & 2.16      & -     & -          & Opt     & 16.1 & 15.5 & 12.0                         & $>$22.9 & s & PS  \\
\hline
\multicolumn{6}{l}{(c) Objects with spectroscopic redshift $z>1$ }\\
\hline
PKS\,0008$-$42                 & 00:10:52.52 & -41:53:10.8 & -         & 6.39      & 1.13  & h         & Opt     & 16.7 & 15.9 & 11.7                          & 22.6     & a  & PS \\
PKS\,0114$-$21                 & 01:16:51.44 & -20:52:06.7 & 4.09      & -         & 1.41  & d          & Opt     & 16.4 & 15.4 & 11.4                         & $>$20.0        & -  & SS \\
PKS\,0402$-$362                & 04:03:53.79 & -36:05:01.7 & 1.15         & 1.03      & 1.423$\dagger$  & g          & WISE     & 12.0 & 10.8 & 7.0                          & 16.7   & c & FS  \\
PKS\,0457+024                & 04:59:52.04 & +02:29:31.6 & 2.07      & -         & 2.384 & i          & WISE     & 15.8  & 14.6 & 11.0                         & 19.0   & c & PS  \\
PKS\,0528$-$250                & 05:30:07.98 & -25:03:30.0 & 1.16      & -         & 2.778 & k$\dagger$ & WISE     & 14.6 & 13.7 & 10.2                          & 17.4   & c & PS  \\
PKS\,0834$-$19                 & 08:37:11.11 & -19:51:56.8 & 4.74      & -         & 1.032 & l          & Opt     & 15.3 & 14.1 & 10.6                         & $>$20.0        & - & PS  \\
4C\,+03.30                   & 14:36:57.18 & +03:24:11.6 & 2.81      & -         & 1.438 & m          & Opt     & 16.5 & 16.1 & 11.8                         & $>$20.0        & - & SS  \\
4C\,+04.51                   & 15:21:14.42 & +04:30:21.7 & 3.93      & -         & 1.296 & n          & Opt     & 15.9 & 14.7 & 11.1                         & 22.2     & o  & FS \\
4C\,+01.69                   & 22:12:37.97 & +01:52:51.3 & 2.80      & -         & 1.126 & q          & Opt     & 15.3 & 14.0  & 10.8                         & 22.0     & q & FS  \\
{[}HB89{]}\,2329$-$162 & 23:31:38.65 & -15:56:57.0 & 1.34      & -         & 1.153 & r          & Opt     & 14.4 & 13.3 & 10.0                         & 20.2   & c  & SS \\
PKS\,2337$-$334                & 23:39:54.54 & -33:10:16.7 & 1.28         & 1.70      & 1.802 & t          & Opt     & 15.4 & 14.3 & 11.6                         & $>$20.0        & - & SS  \\
\hline
\end{tabular}\\
}
{~\\ * {Photometric redshift}\\
$\dagger$ {Literature of non-HI absorption line detected along line of sight with redshift between 0.4~$<$~$z$~$<$~1}\\
$\ddagger$ {Previously searched for H{\sc i} absorption in the literature}\\
Spectroscopic redshift/R-band magnitude reference:
      $^{a}$ {\citet{Burgess2006}},
      $^{b}$ {\citet{deVries1995}},
      $^{c}$ {\citet{Hambly2001}}, 
      $^{d}$ {\citet{McCarthy1996}},
      $^{e}$ {\citet{Tadhunter1993}},
      $^{f}$ {\citet{Holt2008}},
      $^{g}$ {\citet{Jones2009}},
      $^{h}$ {\citet{Labiano2007}},
      $^{i}$ {\citet{Hewitt1986}}, 
      $^{j}$ {\citet{Carilli1998}}, 
      $^{k}$ {\citet{Ellison2010}},
      $^{l}$ {\citet{diSerego-Alighieri1994}},
      $^{m}$ {\citet{Best1999}},
      $^{n}$ {\citet{Heckman1994}},
      $^{o}$ {\citet{diSerego-Alighieri1994}},
      $^{p}$ {\citet{Costa2001}},
      $^{q}$ {\citet{deVries2007}},
      $^{r}$ {\citet{Wright1983}},
      $^{s}$ {\citet{Healey2008}},
      $^{t}$ {\citet{Jackson2002}},
      $^{u}$ {\citet{Snellen2002}}
}
\end{table*}

The aim of our observations is to search for cold neutral gas along the line of sight to radio-loud obscured quasars. All sources were selected from the K\"{u}hr catalogue \citep{Kuhr1981}, a sample of 518 extragalactic radio sources with flux densities $S$~$>$~1~Jy at 5~GHz, most of which were compact. We chose this survey to select bright sources. All sources were required to lie south of declination +5$\degree$ due to the observation limits of ASKAP-BETA. To target sources that were bright at the frequencies we used with the ASKAP-BETA prototype array (700-1000 MHz), we selected sources with a radio flux of 1~Jy or greater in the Sydney University Molonglo Sky Survey \cite[SUMSS;][]{Mauch2003} at 0.843~GHz, or the NRAO VLA Sky Survey \cite[NVSS][]{Condon1998} at 1.4~GHz. Finally, radio sources with known redshifts less than $z$~=~0.4 were discarded, as these are in front of our search volume. 

Two sub-samples were then defined. One was by selecting optically faint AGN as motivated by (and to compare results with) \cite{Carilli1998}, and another selected quasars with red WISE mid-infrared colours (see following sub-sections). A total of 34 radio-bright sources were chosen (Table~\ref{tab:samplesum}). Table~\ref{tab:sourcelist} lists the sources selected and observed for H{\sc i} absorption with ASKAP-BETA, separated by their redshift information. 

\subsubsection{Optically selected sub-sample}\label{sec:optsample}

The first sub-sample features optically faint radio AGN. We cross-matched sources from \cite{Kuhr1981} with the SuperCOSMOS survey \citep{Hambly2001} by search radius (up to 10~arcsec). As in \cite{Carilli1998}, sources with a magnitude of 20~mag or fainter in the R-band were chosen. We also included sources with no R-band SuperCOSMOS optical information, as these sources are likely fainter than the survey sensitivity. We reviewed the literature to remove any sources found with brighter optical magnitudes than 20~mag in other surveys.

In total, this sub-sample has 28 sources, with 15 known to lie between 0.4~$<$~$z$~$<$~1 (can be searched for associated and intervening absorption) from reliable spectroscopic (8) or photometric redshifts (7), from the NASA/IPAC Extragalactic Database (NED). There were 7~optically faint sources which had no known redshift information. These were included in the sample as lower priority targets. We also observed sources from this sub-sample which had known redshifts greater than $z$~=~1 to search for intervening H{\sc i} absorption. This includes PKS\,0008$-$42, an extreme steep spectrum source \citep{Callingham2015}, which had a redshift measurement just above $z$~=~1 \citep{Labiano2007}. This source has two observations carried out at a lower frequency to the rest of the sample (see Table~\ref{tab:obslist}) which enabled a search for associated H{\sc i} absorption for PKS\,0008$-$42. 

\subsubsection{Infrared selected sub-sample}\label{sec:wisesample}

\begin{figure}
\includegraphics[width=1.0\linewidth]{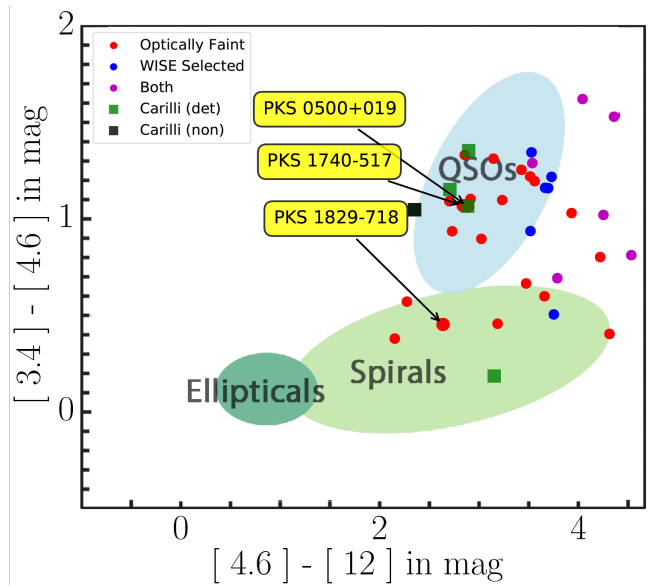}
\caption{WISE two-colour distribution of the sample. The coloured ellipses indicate the typical colour properties seen for elliptical galaxies, spiral galaxies and QSOs, and are reproduced based on fig. 10 of \citet{Wright2010}.} Red circles are sources solely in the optically faint subgroup, blue squares those which were WISE selected, and magenta diamonds objects that fell into both sub-samples. We also include the sample of \citet{Carilli1998} as green (H{\sc i} absorption detections) or black (non-detections) squares for comparison. The three labelled points in our sample (one overlapping with the \citet{Carilli1998} sample) were found to have H{\sc i} absorption toward them (Section~\ref{sec:detections}). Our three detections were all selected from the \citet{Kuhr1981} sample and targeted due to their optical faintness.
\label{fig:wisecolours}
\end{figure}

\begin{figure*}
\centering
\begin{subfigure}[b]{0.49\textwidth}
  \includegraphics[width=1.\linewidth]{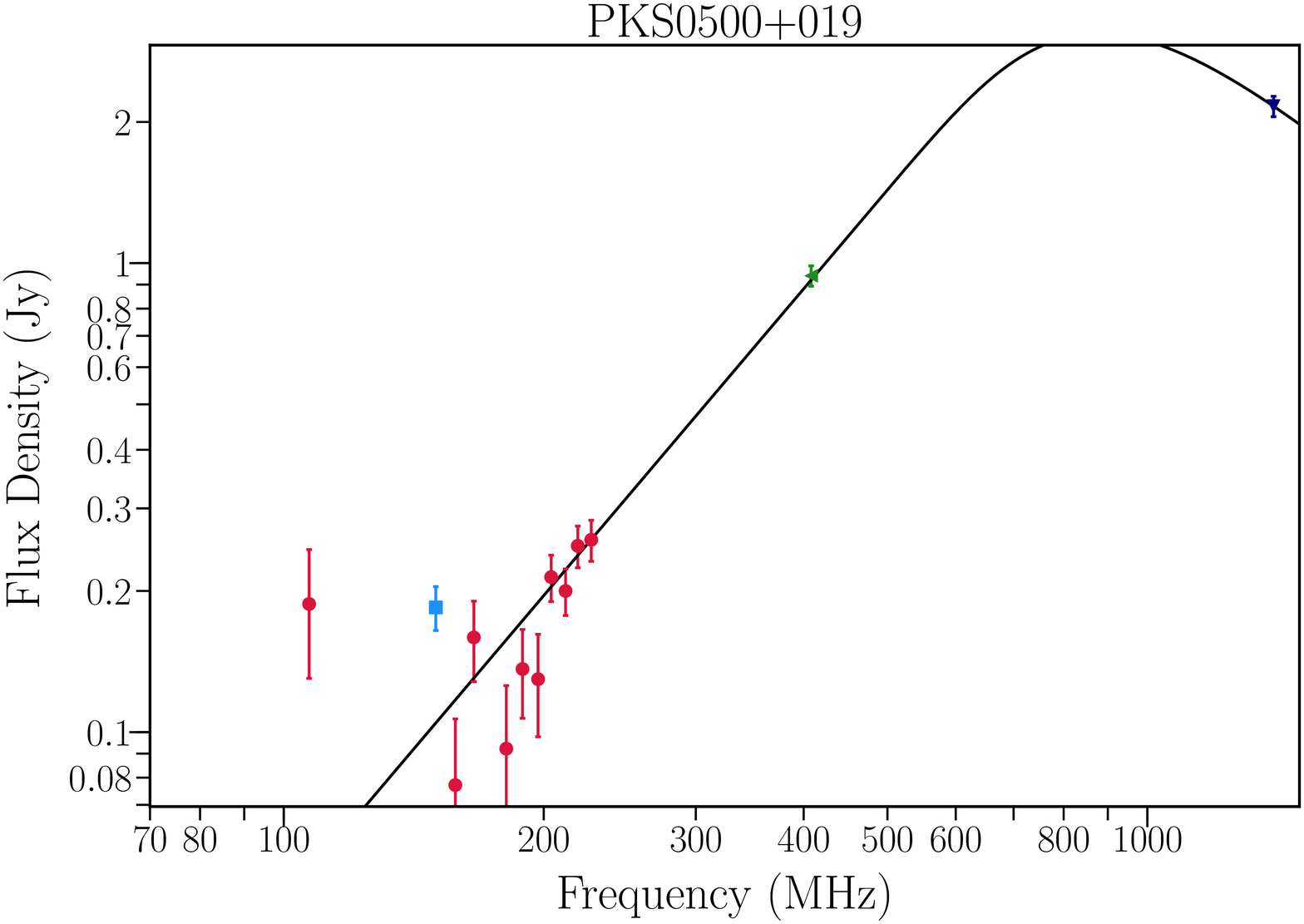}
\end{subfigure}%
\begin{subfigure}[b]{0.49\textwidth}
  \includegraphics[width=1.\linewidth]{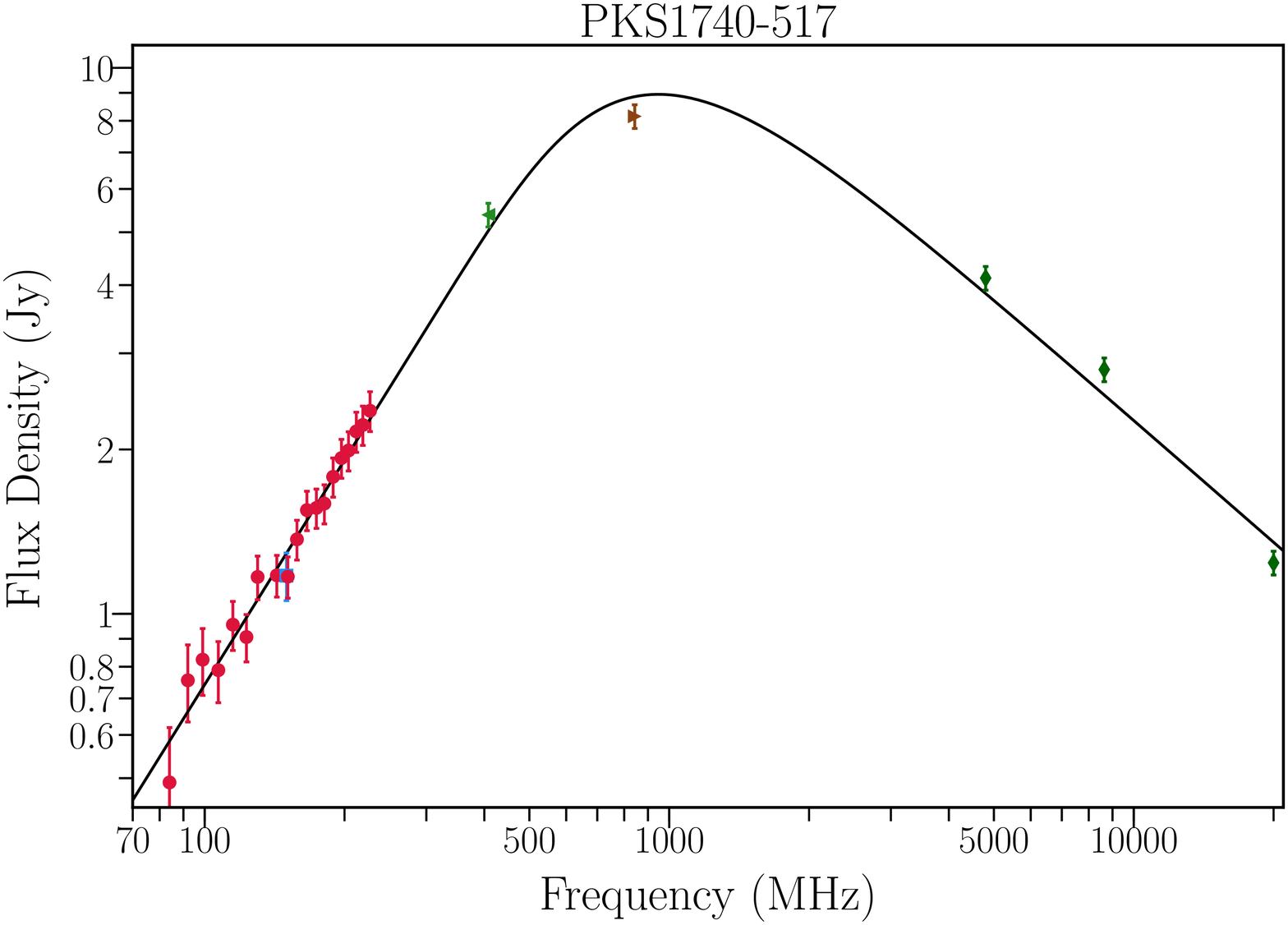}
\end{subfigure}
\begin{subfigure}[b]{0.49\textwidth}
  \includegraphics[width=1.\linewidth]{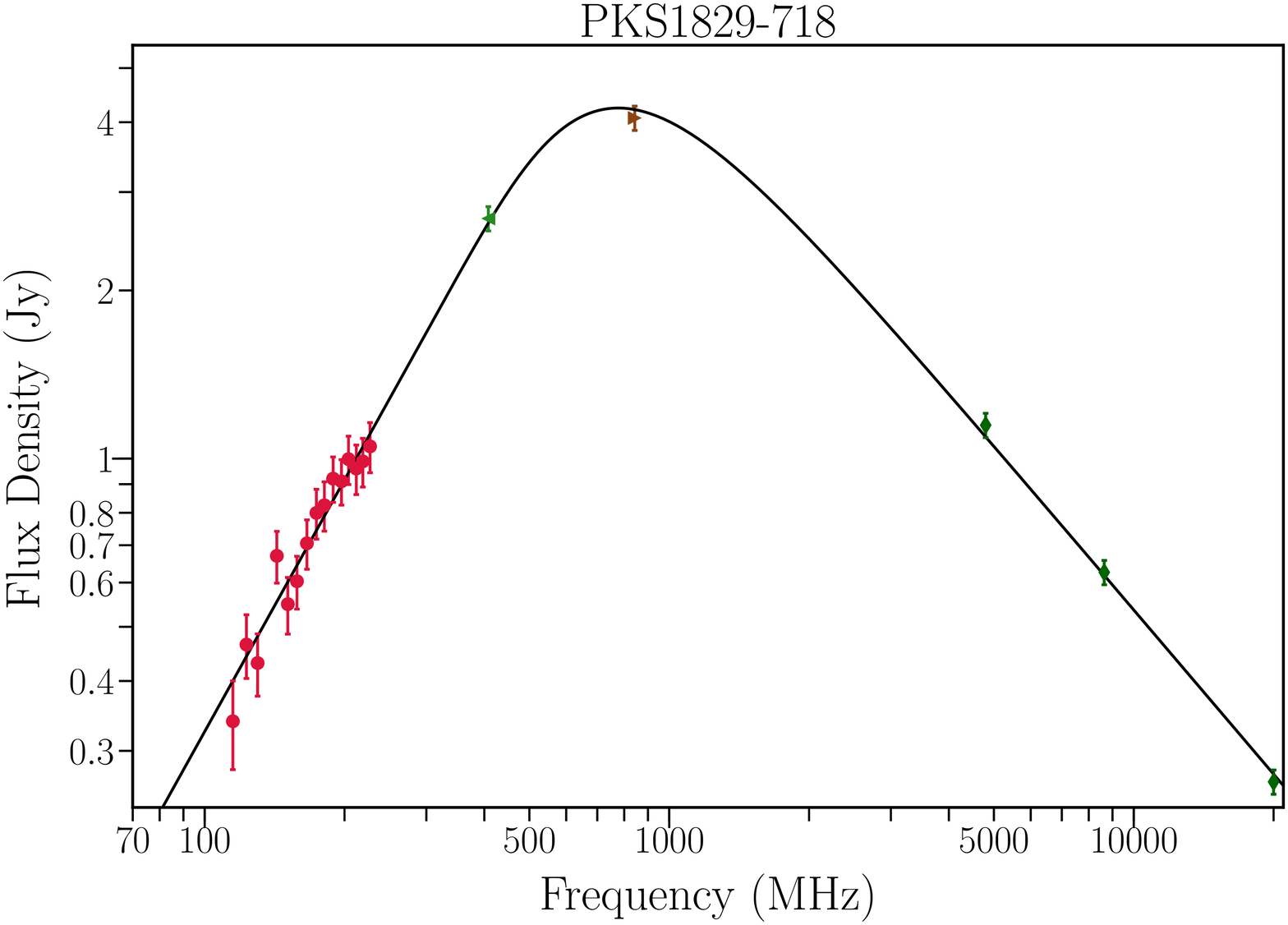}
\end{subfigure}
\caption{Radio spectral energy distributions for three of the peaked spectrum sources within the sample: PKS\,0500+019, PKS\,1740$-$517, and PKS\,1829$-$718. Frequencies range from 72~MHz from the GaLactic and Extragalactic All-sky Murchison Widefield Array (MWA) Survey (GLEAM) survey \citep{Hurley-Walker2017} to 20~GHz from the Australia Telescope 20 GHz (AT20G) survey \citep{Murphy2010} where available for the sources. A generic curved model fit (black) and a power-law fit is made for each SED as per the model fitting code described in \citet{Callingham2015}. All three sources here are identified as peak-spectrum sources by \citet{Callingham2017} and so only have curved model fits present.}
\label{fig:sed}
\end{figure*}

The second sub-sample was selected using data from the Widefield Infrared Survey Explorer \cite[WISE;][]{Wright2010} catalogue. We selected objects with WISE colour $W2-W3$ (4.6~--~12 $\mu$m) of 3.5~mag or greater (Fig.~\ref{fig:wisecolours}; blue or magenta points) to target potentially dustier objects. One motivation is that with redder $W2-W3$ colour we would select objects containing polycyclic aromatic hydrocarbons (PAHs) traced by the W3 band. PAHs are found in star forming regions, potentially containing cold neutral gas we could detect through H{\sc i} absorption.

We required these sources to have a known redshift 0.4~$<$~$z$~$<$~1, either of the host radio galaxy, or an optical absorption line along the line of sight indicating an intervening galaxy that could be reddening the radio host galaxy. Sources in the optical sub-sample without this redshift requirement were only included as lower priority targets. We prioritised observing those with a $W1-W2$ (3.4~--~4.6 $\mu$m) colour of 0.8~mag or greater \cite[i.e. more QSO-like objects;][see Table~\ref{tab:sourcelist}]{Stern2012}. We note that six objects selected as part of the optically faint sub-sample (red points of Fig.~\ref{fig:wisecolours}) would have fallen into this group. However, these six sources had either no spectroscopic or photometric redshifts in the literature between 0.4~$<$~$z$~$<$~1, or did not have an intervening optical absorption line within this redshift range. 

We select 13 sources by their mid-infrared colour information through the criteria described above; here 6 had spectroscopic redshifts 0.4~$<$~$z$~$<$~1, and 5 had only photometric redshift estimates for this redshift range \citep{Burgess2006b}. An overlap with the optically faint sub-sample occurred for 7 sources. Information for the sources are given in Table~\ref{tab:sourcelist}. Compared to the optically faint sub-sample, the WISE-selected radio galaxies overall have similar $W1-W2$ colours (at values where sources in our sample are likely to have a dust-rich circumnuclear region), but redder $W2-W3$ colour (Fig.~\ref{fig:wisecolours}). These galaxies may have higher star formation rates than the optically selected sub-sample.

\subsection{Radio properties of the sample}\label{sec:sed}


To study the SEDs of our sample, we incorporated low-frequency information from the GaLactic and Extragalactic All-sky Murchison Widefield Array (GLEAM) survey \citep{Hurley-Walker2017} to search for turnovers below 1~GHz (i.e. lower-frequency GPS sources). In our sample 12 sources were found to be peaked spectrum (PS) sources (Table~\ref{tab:sourcelist}). Follow-up at VLBI resolutions is required to confirm how compact the PS radio sources in our sample are. The remainder of our full sample were found to be either steep spectrum ($\alpha~>~0.5$), or to have a flat spectral energy distribution (FS; $|\alpha|~<~0.5$) in modelling (4 sources). In Fig.~\ref{fig:sed} we give three example SEDs and fits for sources later discussed in Section~\ref{sec:4}. We also note that three of the four H{\sc i} absorption detections towards reddened quasars in \cite{Carilli1998} were made in GPS radio sources.

\subsection{Photometric redshift estimates}\label{sec:estimate}

\begin{figure*}
\includegraphics[width=1.0\linewidth]{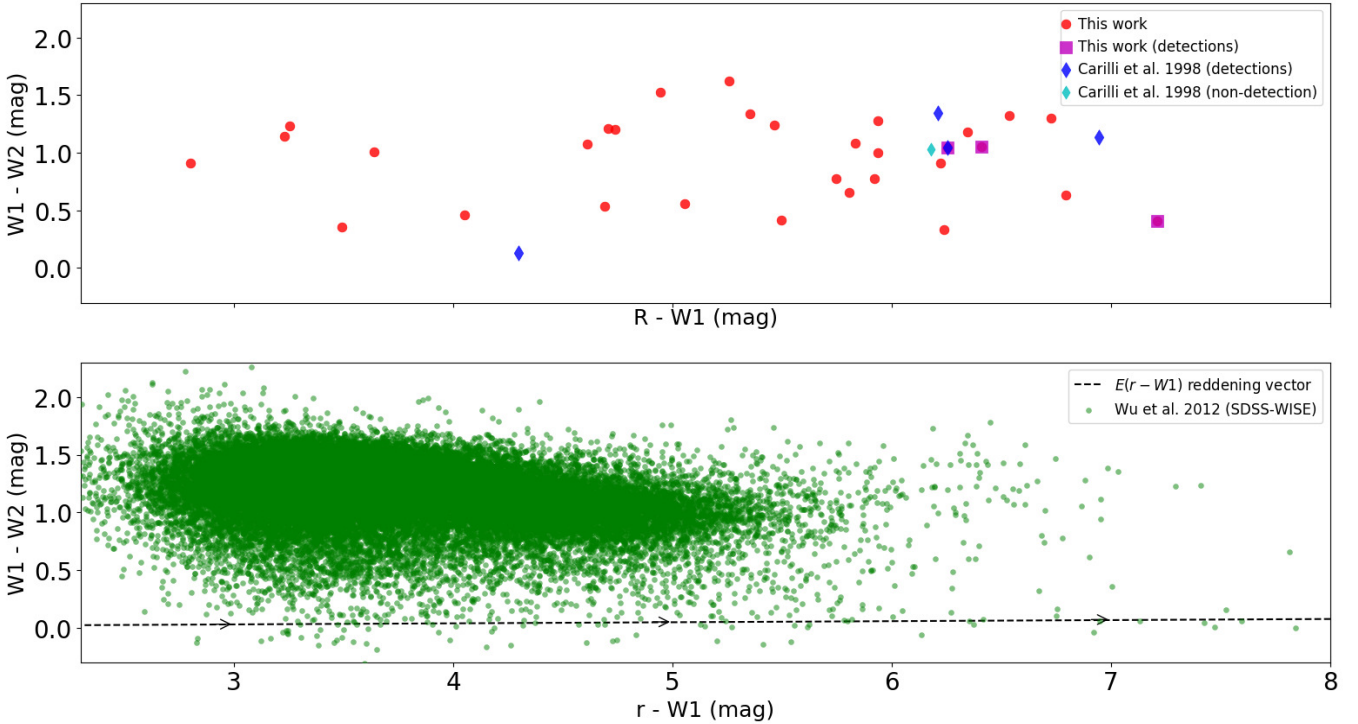}
\caption{Top: Colour-colour plot for our sample and that of \citet{Carilli1998}. The y-axis gives the $W1-W2$ colour from WISE, and the x-axis the optical-to-NIR colour ($R-W1$). Bottom: the SDSS-WISE sample presented in \citet{Wu2012}, with the r-band from SDSS used in place of R from SuperCOSMOS. The $r-W1$ reddening vector assuming $A(V)$ = 1 and $R(V)$~=~3.1 is plotted, as adapted from \citet{Yuan2013}. We can see that the majority of the sources in \citet{Carilli1998} and our sample are significantly redder than most of the SDSS-WISE quasars.}
\label{fig:iroptcolour}
\end{figure*}

In our full sample, 12 sources lack spectroscopic redshifts in the literature, of which 6 had photometric redshift estimates (middle section of Table~\ref{tab:sourcelist}). Additionally, the optical spectra observed of PKS\,0008$-$42 in \cite{Labiano2007} was described as noisy, and hence difficult to distinguish emission lines. This study reported a redshift of $z$~=~1.13 based on [Ne\,{\sc v}]$\lambda$3425 and [O\,{\sc ii}]$\lambda$3727, consistent with photometric redshift predictions \citep{Burgess2006b}. All of these sources were within the optically faint sub-sample. 

To determine if these sources were in front of our search volume, we used R-band redshift estimates, either from \cite{Burgess2006b} or by using the R-$z$ relation provided by that study. As an additional check, we generated redshift probability distributions as a redshift indicator from their available WISE information, as described in \cite{Glowacki2017b}. 

Only PKS\,1213$-$17 was not detected in either WISE or the R-band for these sources, due to proximity to a star. If PKS\,1213$-$17 is infrared-faint, it is likely to lie at high redshift \citep{Norris2006,Collier2014}, but we cannot currently estimate a reliable photometric redshift for this source. All of the remaining sources were estimated to lie at $z$~$>$~0.4; that is, at sufficiently high redshift to search for H{\sc i} absorption with ASKAP. 

\subsection{Dust extinction properties}\label{sec:iroptcolour}

We analyse the dust extinction information available for our sample, and compare with the five-source sample of \cite{Carilli1998}. This is to verify any analysis and comparisons made between the two surveys (see Section~\ref{sec:5}).

We lack access to optical spectra for the majority of our sources, as well as the \cite{Carilli1998} sample, to investigate extragalactic $E(B-V)$ values via template fitting. We confirmed however that Galactic dust reddening and extinction values for both our survey and the \cite{Carilli1998} sample were similar. These were obtained from the NASA/IPAC Infrared Science Archive (IRSA) online service\footnote{\url{https://irsa.ipac.caltech.edu/applications/DUST/}}. We conclude that Galactic dust is not a significant factor in source reddening for our sample.
 
We next compare the optical to near-infrared (NIR) colour information. In Fig.~\ref{fig:iroptcolour} (top panel) we plot the $W1-W2$ colour from WISE versus the $R-W1$ colour for both samples, using values given in Table~\ref{tab:sourcelist}. The $W2-W3$ colour is also given in Fig.~\ref{fig:wisecolours}. Noting the small numbers available for the \cite{Carilli1998} sample, in WISE colour-colour space we see 4 out of 5 \cite{Carilli1998} sources fall in the QSO quadrant (Fig.~\ref{fig:wisecolours}), and all within 2~$<$~$W2-W3$~$<$~3.5~mag. In $R-W1$ colour space, 4 out of 5 \cite{Carilli1998} sources have optical-to-NIR colour~$>$~6~mag, with the full sample spanning 4--7~mag. 

By comparison, our sample spans a wider colour range, although the majority of sources do fall within the full range spanned by all \cite{Carilli1998} sources in Fig.~\ref{fig:iroptcolour} with only a few outliers. In $W1-W2$ space, only two sources in our sample have a higher value than the maximum seen, and none below the lowest seen, in the sample of \cite{Carilli1998}. However, we note that their source with the lowest $W1-W2$ magnitude is an outlier for their sample, and that the significantly smaller \cite{Carilli1998} sample does restrict our ability to fully compare the two distributions. 

In the second panel we compare the two samples with the Sloan Digital Sky Survey (SDSS)-WISE sample \citep{Wu2012}. With the latter, the Sloan r-band is used for the optical-to-NIR colour. While $W1-W2$ colours are similar, in the optical-NIR colour our sample is much redder. From the values given in \cite{Yuan2013}, in substituting $r$ and $W1$ for $B$ and $V$, we find that the extinction for $R-W1$ colours is 2.31 times that of $B-V$ colour. The reddening vector given in the bottom panel highlights that most of our sample is redder than the majority of the \cite{Wu2012} sample. We conclude based on the $R-W1$ colour distribution that, like the \cite{Carilli1998} sample, the majority of our sources are likely reddened by dust in intervening extragalactic systems or in the host radio galaxy, rather than by other means. We discuss the optical-to-NIR properties further in Section~\ref{sec:iroptcolour} in regards to H{\sc i} detection rates of the two surveys.

\begin{table*}
\centering
\caption{Details of the ASKAP-BETA observations. The average channel width is 18.5~kHz. The `Total' row describes the final co-added spectrum for each source. We give the source name, scheduling block identification number (SBID), date, integration time on source, antennas used, the median RMS noise per channel, and the corresponding 1$\sigma$ optical depth sensitivity. Details for PKS\,1740$-$517 and PKS\,2008$-$068 are in \citet{Allison2015} and \citet{Moss2017} respectively. Two observations for PKS\,0008$-$42 (indicated with a~$\ast$) were undertaken in Band 0 (frequency range 657.5--951~MHz). PKS\,0500+019 (indicated with a~$\dagger$) was observed with ASKAP-12.}
\label{tab:obslist}
\begin{tabular}{lrrrccr}
\hline
Source                 & SBID & Date         & $t_{\mathrm{int}}$ & Antenna numbers     & $\sigma_{\mathrm{chan}}$ & $\sigma_{\tau}$      \\
                       &       &              & hr        &                     & mJy beam$^{-1}$ & \%         \\ \hline    
PKS\,0008$-$42           & 2150  & 17 Jul 2015  & 3         & 1,3,6,8,15          & 51              & 1.8                  \\
                       & 2275  & 1 Aug 2015   & 3         & 1,3,6,8,15          & 28              & 3.8                  \\
                       & 3121  & 14 Nov 2015* & 3         & 1,3,6,8,15          & 49              & 0.7                  \\
                       & 3133  & 15 Nov 2015* & 3         & 1,3,6,8,15          & 27              & 0.4                  \\
                       &       & Total        & 12        &                     & 19              & 0.3         \\ \hline    
PKS\,0022$-$423          & 1070  & 11 Nov 2014  & 3.5       & 1,3,6,8,9,15        & 22              & 1.0                  \\
                       & 2680  & 1 Oct 2015   & 4         & 1,3,6,8,15          & 27              & 1.4                  \\
                       & 2685  & 3 Oct 2015   & 5         & 1,3,6,8,15          & 42              & 1.8                  \\
                       &       & Total        & 12.5      &                     & 17              & 0.8           \\ \hline    
PKS\,0042$-$35           & 634   & 15 Sep 2014  & 3         & 1,3,6,8,9,15        & 27              & 0.7                  \\
                       & 3506  & 04 Feb 2016  & 2         & 1,3,8,15            & 100             & 2.6                  \\
                       & 3573  & 13 Feb 2016  & 3         & 1,3,8,15            & 110             & 2.9                  \\
                       &       & Total        & 8         &                     & 19              & 0.5           \\ \hline    
PKS\,0114$-$21           & 2165  & 18 Jul 2015  & 4         & 1,3,6,8,15          & 32              & 0.5                  \\
                       & 2289  & 3 Aug 2015   & 3.5       & 1,3,6,8,15          & 28              & 0.4                  \\
                       &       & Total        & 7.5       &                     & 20              & 0.3           \\ \hline    
3C\,038                  & 559   & 5 Sep 2014   & 3         & 1,3,6,8,9,15        & 53              & 0.7                  \\
                       & 2644  & 28 Sep 2015  & 3         & 1,3,6,8,15          & 33              & 0.5                  \\
                       & 2657  & 29 Sep 2015  & 4         & 1,3,6,8,15          & 28              & 0.3                  \\
                       &       & Total        & 10        &                     & 23              & 0.3           \\ \hline    
PKS\,0235$-$19           & 636   & 15 Sep 2014  & 3         & 1,3,6,8,9,15        & 13              & 0.2                  \\
                       &       & Total        & 3         &                     & 13              & 0.2           \\ \hline    
PKS\,0252$-$71           & 1070  & 11 Nov 2014  & 3.5       & 1,3,6,8,9,15        & 23              & 0.2                  \\
                       & 2450  & 25 Aug 2015  & 3         & 1,3,6,8,9,15        & 37              & 0.4                  \\
                       & 2675  & 01 Oct 2015  & 5         & 1,3,6,8,15          & 26              & 0.3                  \\
                       & 2680  & 02 Oct 2015  & 2         & 1,3,6,8,15          & 26              & 0.3                  \\
                       & 3317  & 12 Dec 2015  & 3         & 1,3,6,8,15          & 29              & 0.3                  \\
                       & 3321  & 13 Dec 2015  & 6         & 1,3,6,8,15          & 20              & 0.2                  \\
                       &       & Total        & 22.5      &                     & 10              & 0.1           \\ \hline    
PKS\,0402$-$362          & 560   & 5 Sep 2014   & 3         & 1,3,6,8,9,15        & 34              & 3.6                  \\
                       & 3506  & 4 Feb 2016   & 3         & 1,3,8,15            & 100             & 10.6                 \\
                       & 3573  & 13 Feb 2016  & 3         & 1,3,8,15            & 38              & 4.7                  \\
                       &       & Total        & 9         &                     & 25              & 2.8           \\ \hline    
PKS\,0408$-$65           & 1096  & 14 Nov 2014  & 3         & 1,3,6,8,9,15        & 29              & 0.1                  \\
                       &       & Total        & 3         &                     & 29              & 0.1           \\ \hline    
PKS\,0420-01           & 1528  & 11 Mar 2015  & 2         & 1,3,8,9,15          & 48              & 1.7                  \\
                       &       & Total        & 2         &                     & 48              & 1.7           \\ \hline    
3C\,118                  & 2077  & 7 July 2015  & 3         & 1,3,6,8,15          & 33              & 1.7                  \\
                       & 3176  & 21 Nov 2015  & 2.5       & 1,3,6,8,15          & 46              & 1.8                  \\
                       & 3181  & 22 Nov 2015  & 3         & 1,3,6,8,15          & 52              & 2.3                  \\
                       &       & Total        & 8.5       &                     & 22              & 1.0           \\ \hline    
PKS\,0457+024          & 1767  & 6 May 2015   & 2         & 1,3,6,8,9,15        & 32              & 2.6                  \\
                       &       & Total        & 2         &                     & 32              & 2.6           \\ \hline    
PKS\,0500+019$\dagger$ & 3348  & 4 Feb 2017   & 2         & 2,3,4,5,6,10,12,13,14, & 14              & 0.7                  \\
 &   &    &          & 19,24,26,27,28  &               &                  \\
                       &       & Total        & 2         &   & 14              & 0.7           \\ \hline    
PKS\,0528$-$250          & 1096  & 14 Nov 2014  & 3         & 1,3,6,8,9,15        & 27              & 3.3                  \\
                       &       & Total        & 3         &                     & 27              & 3.3           \\ \hline    
   \end{tabular}
\end{table*}
\begin{table*}
\ContinuedFloat
\centering
\caption{Continued.}
\begin{tabular}{lrrrccr}
\hline
Source                 & SBID & Date         & $t_{\mathrm{int}}$ & Antenna numbers     & $\sigma_{\mathrm{chan}}$ & $\sigma_{\tau}$     \\
                       &       &              & hr        &                     & mJy beam$^{-1}$ & \%         \\ \hline    
PKS\,0834$-$19           & 2146  & 15 Aug 2015  & 3         & 1,3,6,8,15          & 34              & 0.5                  \\
                       & 3317  & 12 Dec 2015  & 3         & 1,3,6,8,15          & 33              & 0.4                  \\
                       & 3321  & 13 Dec 2015  & 3         & 1,3,6,8,15          & 34              & 0.5                  \\
                       &       & Total        & 9         &                     & 22              & 0.3           \\ \hline    
PKS\,1213$-$17           & 1999  & 26 Jun 2015  & 3         & 1,3,6,8,15          & 37              & 2.1                  \\
                       & 3574  & 13 Feb 2016  & 3         & 1,3,8,15            & 43              & 2.0                  \\
                       &       & Total        & 6         &                     & 23              & 1.2           \\ \hline    
3C\,275                  & 1147  & 22 Nov 2014  & 2         & 1,3,6,8,9,15        & 32              & 0.5                  \\
                       & 1150  & 23 Nov 2014  & 2         & 1,3,6,8,9,15        & 32              & 0.5                  \\
                       &       & Total        & 4         &                     & 19              & 0.3           \\ \hline    
4C\,+03.30               & 2266  & 31 Jul 2015  & 3         & 1,3,6,8,15          & 30              & 0.8                  \\
                       &       & Total        & 3         &                     & 30              & 0.8           \\ \hline    
4C\,+04.51               & 3500  & 25 Jul 2015  & 5         & 1,3,8,15            & 24              & 0.6                  \\
                       &       & Total        & 5         &                     & 24              & 0.6           \\ \hline    
PKS\,1622$-$253          & 2640  & 28 Sep 2015  & 3         & 1,3,6,8,15          & 30              & 1.7                  \\
                       & 3178  & 22 Nov 2015  & 3         & 1,3,6,8,15          & 44              & 2.0                  \\
                       & 3318  & 12 Dec 2015  & 5         & 1,3,6,8,15          & 28              & 1.1                  \\
                       &       & Total        & 11        &                     & 16              & 0.8           \\ \hline    
PKS\,1829$-$718          & 2108  & 12 Jul 2015  & 5         & 1,6,8,15            & 34              & 1.0                  \\
                       & 2674  & 30 Sep 2015  & 6         & 1,3,6,8,15          & 22              & 0.6                  \\
                       & 2695  & 6 Oct 2015   & 4.5       & 1,3,6,8,15          & 27              & 0.7                  \\
                       &       & Total        & 15.5      &                     & 14              & 0.4           \\ \hline    
PKS\,1936$-$623          & 2120  & 13 Jul 2015  & 4         & 1,3,6,8,15          & 37              & 1.8                  \\
                       & 3179  & 22 Nov 2015  & 4         & 1,3,6,8,15          & 32              & 2.0                  \\
                       &       & Total        & 8         &                     & 28              & 1.4           \\ \hline    
PKS\,2032$-$35           & 557   & 5 Sep 2014   & 3         & 1,3,6,8,9,15        & 37              & 0.4                  \\
                       & 1096  & 14 Nov 2014  & 3         & 1,3,6,8,9,15        & 25              & 0.3                  \\
                       &       & Total        & 6         &                     & 18              & 0.2           \\ \hline    
PKS\,2140$-$81           & 2657  & 29 Sep 2015  & 4         & 1,3,6,8,15          & 27              & 2.7                  \\
                       &       & Total        & 4         &                     & 27              & 2.7           \\ \hline    
PKS\,2150$-$52           & 632   & 15 Sep 2014  & 3         & 1,3,6,8,9,15        & 27              & 0.5                  \\
                       & 2503  & 4 Feb 2016   & 2         & 1,3,8,15            & 100             & 1.9                  \\
                       &       & Total        & 5         &                     & 16              & 0.3           \\ \hline    
4C\,+01.69               & 2330  & 8 Aug 2015   & 4.5       & 1,3,6,8,15          & 38              & 1.0                  \\
                       &       & Total        & 4.5       &                     & 38              & 1.0           \\ \hline    
PKS\,2307$-$282          & 2339  & 11 Aug 2015  & 5         & 1,6,8,15            & 29              & 1.7                  \\
                       &       & Total        & 5         &                     & 29              & 1.7           \\ \hline    
PKS\,2323$-$40           & 557   & 5 Sep 2014   & 3         & 1,3,6,8,9,15        & 36              & 0.7                  \\
                       & 1757  & 6 May 2015   & 3         & 1,3,6,8,9,15        & 23              & 0.5                  \\
                       &       & Total        & 6         &                     & 21              & 0.4           \\ \hline    
{[}HB89{]}\,2329$-$162   & 2327  & 7 Aug 2015   & 5         & 1,6,8,15            & 35              & 3.7                  \\
                       &       & Total        & 5         &                     & 35              & 3.7           \\ \hline    
PKS\,2331$-$41           & 633   & 15 Sep 2014  & 3         & 1,3,6,8,9,15        & 29              & 0.3                  \\
                       &       & Total        & 3         &                     & 29              & 0.3           \\ \hline    
PKS\,2333$-$528          & 2718  & 13 Aug 2015  & 4         & 1,3,6,8,9,15        & 29              & 1.4                  \\
                       & 3181  & 22 Nov 2015  & 3         & 1,3,6,8,9,15        & 56              & 2.6                  \\
                       &       & Total        & 7         &                     & 20              & 0.9           \\ \hline    
PKS\,2337$-$334          & 2327  & 7 Aug 2015   & 4         & 1,3,6,8,9,15        & 34              & 2.0                  \\
                       &       & Total        & 4         &                     & 34              & 2.0           \\ \hline
\end{tabular}
\end{table*}

\section{Observations}\label{sec:3}

\subsection{Observations with ASKAP}
We conducted observations with the six-antenna prototype array of ASKAP (ASKAP-BETA) from September 2014 to February 2016, between 712.5--1015.5~MHz. We chose this frequency band of ASKAP due to the corresponding redshift range (0.4~$<$~$z$~$<$~1). Observations were carried out in minimum 2~hour blocks for each source.  Table~\ref{tab:obslist} describes the observing dates, antennas used and integration time for each source, as well as the optical depth sensitivity reached from the combined observations. Towards the end of the observing period only a subset of the six antennas was available for observation, which lowered the sensitivity achieved relative to observations undertaken with all six antennas. 

The data were reduced following the procedure described by \cite{Allison2015}. For each individual source with multiple observations, all good datasets were combined to produce a final spectrum. When considering the root mean square (RMS) noise for each final combined spectrum for each target, the median achieved is 22.5~mJy\,beam$^{-1}$; see Table~\ref{tab:obslist} for the RMS noise achieved for each source. The average channel width is 18.5~kHz, equivalent to H{\sc i} velocities of 5.5--7.8~km\,s$^{-1}$. 
Many sources required re-observation due to commissioning issues encountered (e.g. correlator block failures); such errors resulted in gaps in the spectrum. We undertook longer observations for some sources to improve on areas of the bandwidth which were corrupted. Sources such as PKS\,2337$-$334 were afflicted by correlator block errors in segments of the combined spectra. For instance, 16 of the final combined spectra had corrupted data due to a correlator block error centred at $\sim$990~MHz. We estimate about 2\% of the total redshift space was corrupted, and hence cannot be searched for H{\sc i} absorption. 

Our observations for PKS\,0008$-$42 ($z$~=~1.13) includes two `Band~0' observations, extending down to 657.5~MHz. This corresponds to a redshift range for which we could detect H{\sc i} absorption of 0.4~$<$~$z$~$<$~1.17, allowing a search for associated H{\sc i} absorption for this source. PKS\,0500+019 had an ASKAP-12 observation taken in February 2017, motivated due to a correlator block failure at the redshift of the known H{\sc i} absorption feature during the ASKAP-BETA observation. ASKAP-12, which was the next development stage for the ASKAP commissioning telescope, featured at the time 14 antennas fitted with the Mk-II Phased Array Feeds (PAFs) which had lower system temperatures to the Mk-I PAFs \citep{Chippendale2015}.

While PKS\,0500+019 was the only source whose ASKAP-BETA observations had corrupted chunks at the known spectroscopic redshift for the host radio galaxy, correlator block errors seen in other observations may have prevented the detection of potential intervening H{\sc i} absorption. Correlator block errors also negatively affect our H{\sc i} absorption search towards sources for which we had no spectroscopic redshift information.

\begin{figure*}
\includegraphics[width=1.0\linewidth]{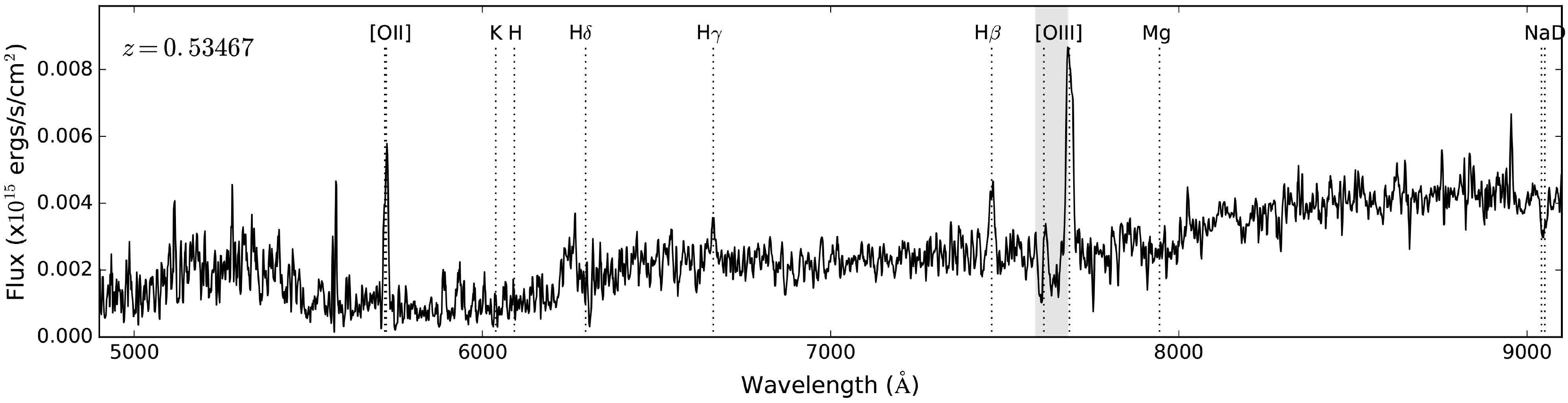}
\caption{Optical spectrum taken with GMOS-S on Gemini South of PKS\,1829$-$718. We observe strong [O\,{\sc ii}], [O\,{\sc iii}] and H$\beta$ emission lines, identifying the galaxy as a high-excitation radio galaxy. We find a redshift of $z$~=~0.53467 from fitting with a high redshift star forming galaxy template within the MARZ program \citep{Hinton2016}. This confirms that the H{\sc i} absorption we detect is associated with the host galaxy of the radio AGN. The gray shaded area indicates the location of an atmospheric absorption band, which here overlaps with the [O\,{\sc iii}] line. Some residual night sky lines are seen in the spectrum, such as at 5,577 angstroms.}
\label{fig:opticalspectrum}
\end{figure*}

\begin{figure}
\includegraphics[width=1.0\linewidth]{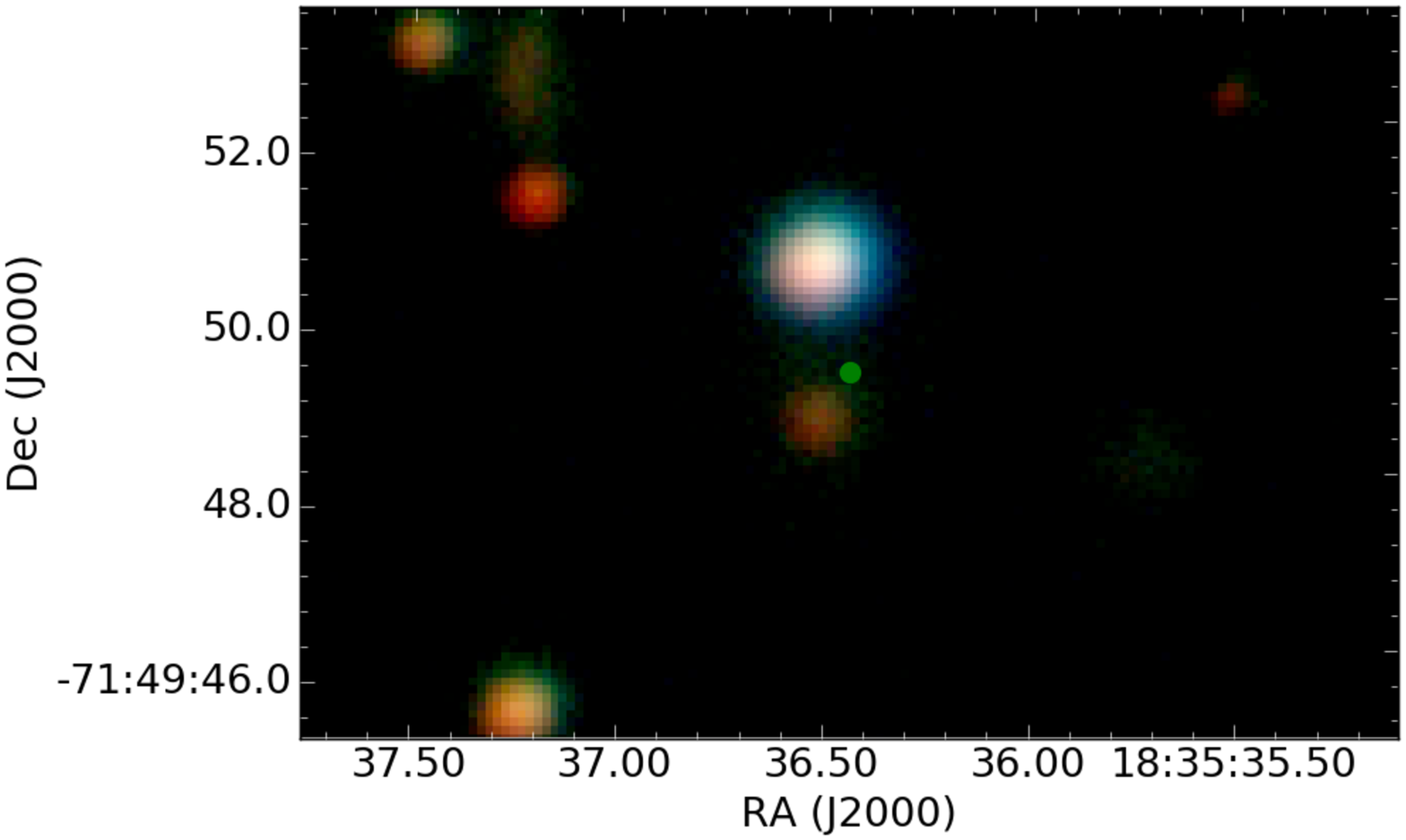}
\caption{Optical three-colour image constructed from the $g'$ (blue), $r'$ (green), and $i'$ (red) band observations with GMOS-S on Gemini South of PKS\,1829$-$718. The green circle gives the centre of the radio position from the literature (NED). The central radio position is slightly offset ($\sim$1~arcsec) with a faint red optical source we believe is associated, from which we extracted an optical spectrum (Fig.~\ref{fig:opticalspectrum}). We note that the radio source is unresolved both in our ASKAP-BETA observation and the literature. Error bars are not included as errors for the radio position given in the literature are too small to represent ($<$0.01 seconds and arcseconds in RA and Dec respectively.) The red optical source was not seen in previous optical surveys.}
\label{fig:opticalimage}
\end{figure}

We searched the final combined spectra of each source for H{\sc i} absorption features. This was done both through visual inspection, and an automated line finder and parameterisation pipeline, which incorporated Bayesian statistical methods \citep{Allison2012b}. The latter method was used to verify any candidates or detect features, and quantify the H{\sc i} column density measurements and upper values for non-detections. We discuss our detections in Section~\ref{sec:4}. We give the full ASKAP-BETA spectra for PKS\,0008$-$42, PKS\,0500+019, and PKS\,1829$-$718 as separate figures in online supplementary material.


\begin{table*}
\centering
\caption{Derived optical depth and H{\sc i} column density upper limits and values for our sources observed with ASKAP-BETA. We give the total integration time, median flux density measured in the ASKAP band, the observed/upper limit peak optical depth at five times the optical depth sensitivity given in Table~\ref{tab:obslist} (5$\sigma_{\tau}$), velocity integrated optical depth (assuming a FWHM of 30\,km\,s$^{-1}$), H{\sc i} column density multiplied by a covering factor $f$ (assuming a spin temperature of 100\,K) sensitivity, and for detections the median redshift of the absorption. These values are typical of other H{\sc i} absorption detections and studies. Details for PKS\,1740$-$517 and PKS\,2008$-$068 are from \citet{Allison2015} and \citet{Moss2017} respectively.}
\label{tab:derivedvalues}
\begin{tabular}{llccrrrc}
\hline
\# & Source        & t$_{int}$ & S$_{\mathrm{med}}$ & $\tau_{\mathrm{obs,peak}}$ & $\int{\tau_\mathrm{obs}\mathrm{d}v}$ & $N_\mathrm{HI}f$ & $z_{\rm abs}$    \\
   &               & hr        & Jy                 & & km\,s$^{-1}$                         & cm$^{-2}$ &                       \\ \hline
1  & PKS\,0008$-$42  & 12        & 7.03               & $<$0.006 & $<$0.35                              & $<$6.5\,$\times$\,10$^{19}$ & --\\
2  & PKS\,0022$-$423          & 13.5      & 2.07  & $<$0.041                  & $<$1.03                              & $<$1.9\,$\times$\,10$^{20}$ & --\\
3  & PKS\,0042$-$35           & 8         & 3.65     & $<$0.026                  & $<$0.64                              & $<$1.2\,$\times$\,10$^{20}$ & --\\
4  & PKS\,0114$-$21           & 7.5       & 7.14     & $<$0.014                  & $<$0.28                              & $<$5.1\,$\times$\,10$^{19}$ & --\\
5  & 3C\,038                  & 10        & 7.67     & $<$0.015                  & $<$0.46                              & $<$8.5\,$\times$\,10$^{19}$ & --\\
6  & PKS\,0235$-$19           & 3         & 5.91      & $<$0.011                  & $<$0.33                              & $<$6.0\,$\times$\,10$^{19}$ & --\\
7  & PKS\,0252$-$71           & 22.5      & 10.02     & $<$0.005                  & $<$0.15                              & $<$2.8\,$\times$\,10$^{19}$ & --\\
8  & PKS\,0402$-$362          & 9         & 0.91    & $<$0.138                  & $<$3.74                              & $<$6.8\,$\times$\,10$^{20}$ & --\\
9  & PKS\,0408$-$65           & 3         & 29.00    & $<$0.005                  & $<$0.12                              & $<$2.3\,$\times$\,10$^{19}$ & --\\
10 & PKS\,0420-01           & 2         & 2.86   & $<$0.084                  & $<$2.12                              & $<$3.9\,$\times$\,10$^{20}$ & --\\
11 & 3C\,118                  & 8.5       & 2.16      & $<$0.051                  & $<$1.72                              & $<$3.1\,$\times$\,10$^{20}$ & --\\
12 & PKS\,0457+024          & 2         & 1.22      & $<$0.131                  & $<$1.88                              & $<$3.4\,$\times$\,10$^{20}$ & --\\
13 & PKS\,0500+019 & 2         & 1.88       & 0.037                  & 3.7                              & 6.7\,$\times$\,10$^{20}$ & 0.585\\
14 & PKS\,0528$-$250          & 3         & 0.81    & $<$0.166                  & $<$2.13                              & $<$3.9\,$\times$\,10$^{20}$ & --\\
15 & PKS\,0834$-$19           & 12        & 7.33   & $<$0.015                  & $<$0.36                              & $<$6.5\,$\times$\,10$^{19}$ & --\\
16 & PKS\,1213$-$17           & 6         & 1.89    & $<$0.061                  & $<$1.74                              & $<$3.2\,$\times$\,10$^{20}$ & --\\
17 & 3C\,275                  & 4         & 5.94    & $<$0.016                  & $<$0.50                              & $<$9.1\,$\times$\,10$^{19}$ & --\\
18 & 4C\,+03.30               & 3         & 3.66    & $<$0.041                  & $<$0.81                              & $<$1.5\,$\times$\,10$^{20}$ & --\\
19 & 4C\,+04.51               & 5         & 4.29    & $<$0.028                  & $<$0.59                              & $<$1.1\,$\times$\,10$^{20}$ & --\\
20 & PKS\,1622$-$253          & 11        & 2.05    & $<$0.039                  & $<$1.05                              & $<$1.9\,$\times$\,10$^{20}$ & --\\
21 & PKS\,1740$-$517 & 26.4 & 8.15 & 0.204 & 2.7 & 5.0\,$\times$\,10$^{20}$ & 0.441\\
22 & PKS\,1829$-$718          & 15.5      & 3.50   & 0.246                  & 27.4                              & 5.0\,$\times$\,10$^{21}$ & 0.536 \\
23 & PKS\,1936$-$623          & 8         & 1.97    & $<$0.071                  & $<$1.99                              & $<$3.6\,$\times$\,10$^{20}$ & --\\
24 & PKS\,2008$-$068 & 15.5 & 3.04 & $<$0.028 & $<$0.88 & $<$1.6\,$\times$\,10$^{20}$ & --\\
25 & PKS\,2032$-$35           & 6         & 8.18     & $<$0.011                  & $<$0.34                              & $<$6.2\,$\times$\,10$^{19}$ & --\\
26 & PKS\,2140$-$81           & 4         & 3.86      & $<$0.035                  & $<$1.03                              & $<$1.9\,$\times$\,10$^{20}$ & --\\
27 & PKS\,2150$-$52           & 5         & 5.71      & $<$0.014                  & $<$0.38                              & $<$6.9\,$\times$\,10$^{19}$ & --\\
28 & 4C\,+01.69               & 4.5       & 3.80     & $<$0.050                  & $<$1.14                              & $<$2.1\,$\times$\,10$^{20}$ & --\\
29 & PKS\,2307$-$282          & 5         & 1.77     & $<$0.082                  & $<$2.68                              & $<$4.9\,$\times$\,10$^{20}$ & --\\
30 & PKS\,2323$-$40           & 6         & 5.01    & $<$0.021                  & $<$0.56                              & $<$1.0\,$\times$\,10$^{20}$ & --\\
31 & [HB89]\,2329$-$162          & 5         & 1.30      & $<$0.134                  & $<$3.02                              & $<$5.5\,$\times$\,10$^{20}$ & --\\
32 & PKS\,2331$-$41           & 3         & 9.06      & $<$0.016                  & $<$0.41                              & $<$7.4\,$\times$\,10$^{19}$ & --\\
33 & PKS\,2333$-$528          & 7         & 2.17       & $<$0.046                  & $<$1.38                              & $<$2.5\,$\times$\,10$^{20}$ & --\\
34 & PKS\,2337$-$334          & 4         & 1.70       & $<$0.100                  & $<$1.73                              & $<$3.2\,$\times$\,10$^{20}$ & --\\ \hline
\end{tabular}
\end{table*}

\subsection{Follow-up optical spectroscopy of PKS 1829$-$718}\label{sec:opticalspect}

Prior to our observations with ASKAP-BETA, the redshift of PKS\,1829$-$718 was not known. There is no spectroscopic measurement in the literature and only a possible optical identification made at $\sim$23~mag \citep{Costa2001}. This source was detected in H{\sc i} absorption during our observations (Section~\ref{sec:4}). While the photometric redshift indicator method of \cite{Glowacki2017b} predicted that the H{\sc i} absorption we detect is associated with the host galaxy, the redshift probability distribution generated was unreliable due to confusion with a nearby source in WISE. To properly interpret the H{\sc i} absorption, it was necessary to determine if the H{\sc i} gas was associated with the host galaxy of the radio AGN, or in an intervening system along the line of sight.

We obtained long-slit optical spectra on 21 August 2017 with the Gemini Multi-Object Spectrograph-South \cite[GMOS-S;][]{Hook2003} on Gemini South (proposal code GS-2017B-Q-63, PI Ellison). The long-slit spectroscopy was taken with the R400 grating. Two grating settings with central wavelengths at 7000 and 7050~\AA~were used, in order to detect [O\,{\sc ii}], [O\,{\sc iii}] and H$\beta$ in emission. At each grating setting we obtained 4~$\times$~900\,s exposures, giving a total exposure time of 2~hours. A 1.5-arcsec slit was used, and the CCD was binned 2~$\times$~2 on-chip. We followed the standard data reduction processes through the Gemini data reduction packages in PYRAF.

The optical spectrum is shown in Fig.~\ref{fig:opticalspectrum}. Through the Manual and Automatic Redshifting Software program \cite[MARZ;][]{Hinton2016}, we find a best fit with a high redshift star forming galaxy template. We identify strong [O\,{\sc ii}], [O\,{\sc iii}] and H$\beta$ emission lines, typical of high-excitation radio galaxies. Through this template fit we obtain a redshift of $z$~=~0.5347~$\pm$~0.0004, with a maximum cross-correlation value of 10.495 (a measure of the agreement of the template fit with the optical spectrum). We compare this with the H{\sc i} absorption redshift detected in Section~\ref{sec:detections}.

We also acquired images with GMOS in the $g'$, $r'$, and $i'$ bands. Each image consisted of 4~$\times$~75\,s exposures, which were binned 2~$\times$~2 on-chip, giving a pixel scale of 0.16\,arcsec\,pixel$^{-1}$. A three-colour image ($g'$ as blue, $r'$ as green, and $i'$ as red) is shown in Fig.~\ref{fig:opticalimage}. Adjusting the original astrometry of the Gemini image through Astrometry.net \citep{Lang2010}, we find that the radio position of PKS\,1829$-$718 (indicated by the green marker) is close to a faint and red optical galaxy, which was not previously detected in optical surveys. The offset between the radio position and the centre of the optical source position ($\sim$1~arcsec) is greater than the radio position error ($\sim$0.01~arcsec) found in the literature (NED). However, the radio source is unresolved in our observations and the literature, e.g. unresolved to 10 arcseconds in AT20G. Higher spatial resolution observations are required to investigate the extent of the radio source and any offset between it and the optical source we see. 

\section{Results}\label{sec:4}

We detect H{\sc i} absorption towards three sources in our sample. Two were not identified prior to the ASKAP-BETA observations, with one of those (PKS\,1740$-$517) reported in \cite{Allison2015}. The H{\sc i} absorption detection rate for sources we have spectroscopic redshifts available, and searched for associated absorption, is 25\% (3/12), with a standard error of 12.5\%. When extending to sources with photometric redshift estimates consistent with $0.4~<~z~<~1$, this detection rate becomes 12.5\% $\pm$ 6.8\% (3/24). All three detections were part of the optically faint sub-sample. 

\subsection{Individual H{\sc i} absorption detections}\label{sec:detections}
\subsubsection{PKS 1829$-$718}

\begin{figure}
\includegraphics[width=1.0\linewidth]{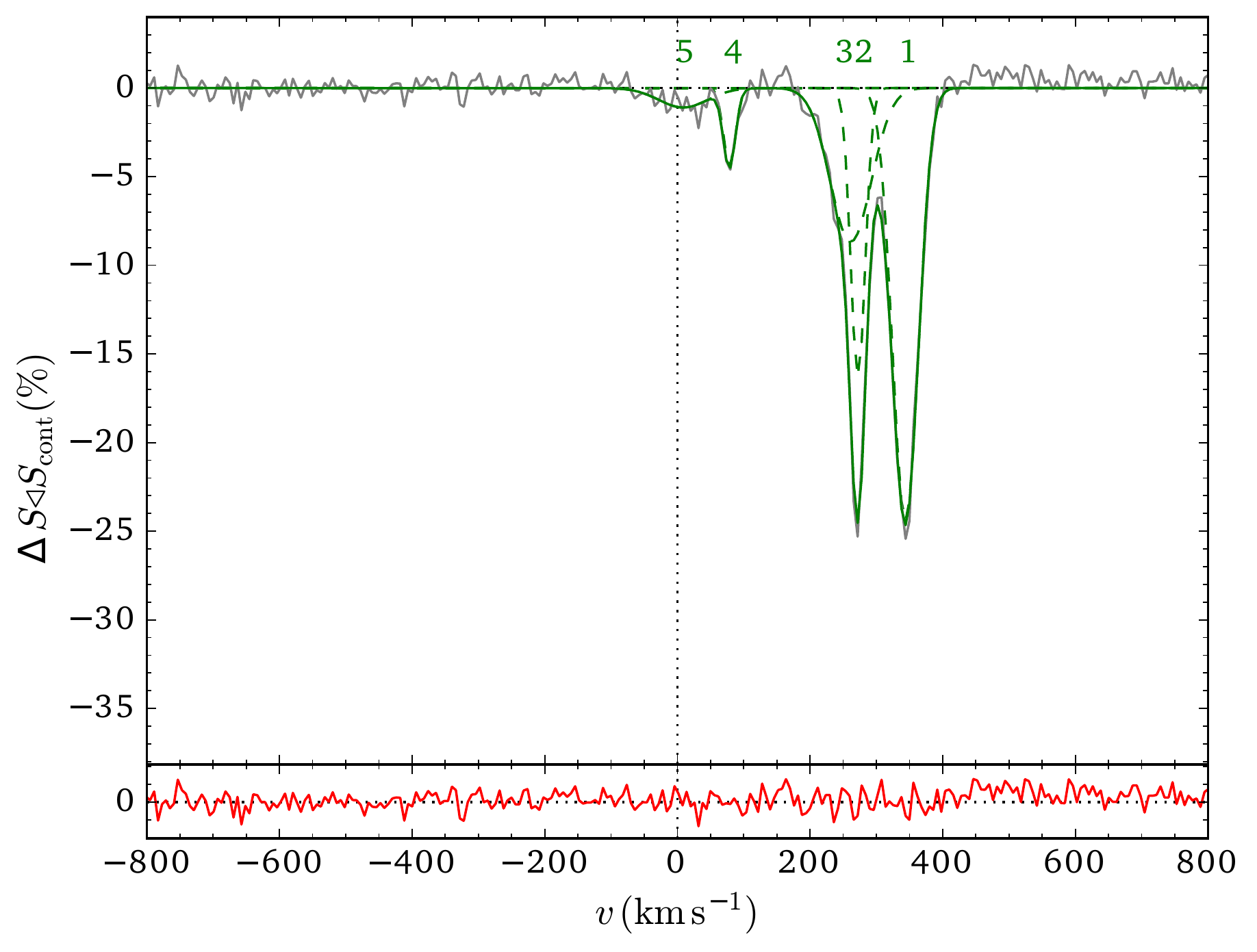}
\caption{The H{\sc i} absorption detection made towards PKS\,1829$-$718. The top panel shows the data following subtraction of the best-fitting continuum model (grey), and the best-fitting multiple Gaussian spectral-line model (black) as given in Table~\ref{tab:analysis_det}. The individual components are displayed in green. The red line in the bottom panel represents the residuals of the data after subtraction of this model fit. The spectrum is centred on the optical spectroscopic redshift obtained of $z$~=~0.53467 through template fitting in MARZ, which has an error of $z$~=~0.0004.}.
\label{fig:PKS1829_fit}
\end{figure}

We detected H{\sc i} absorption towards PKS\,1829$-$718 centred at redshift $z_{\mathrm{abs}}$~=~0.536. We model the H{\sc i} absorption features detected towards PKS\,1829$-$718 as a superposition of Gaussian profiles and determine the best fitting line parameters (number of components and marginal distributions) through a Bayesian approach by comparing against other models, including a no absorption model \citep{Allison2012b}, as done in \cite{Allison2015}. Our best fit is given in Fig.~\ref{fig:PKS1829_fit}. We identify two deep absorption features at a redshift $z$~$\sim$~0.536 best modelled through three Gaussian components, and two shallower blueshifted components (Table~\ref{tab:analysis_det}) at $z$~$\sim$~0.535. The peak and integrated optical depths of the H{\sc i} absorption are $0.249 \pm 0.004$ and $\int \tau {\rm d}v$\,km~s$^{-1}$~=~$27.4 \pm 0.3$, respectively. Assuming a spin temperature of 100\,K and a covering factor of 1 \cite[motivated by the compact size of the radio source;][]{Murphy2010}, we obtain a column density of 5.0~$\times$~10$^{21}$\,cm$^{-2}$. 

\begin{figure}
\includegraphics[width=1.0\linewidth]{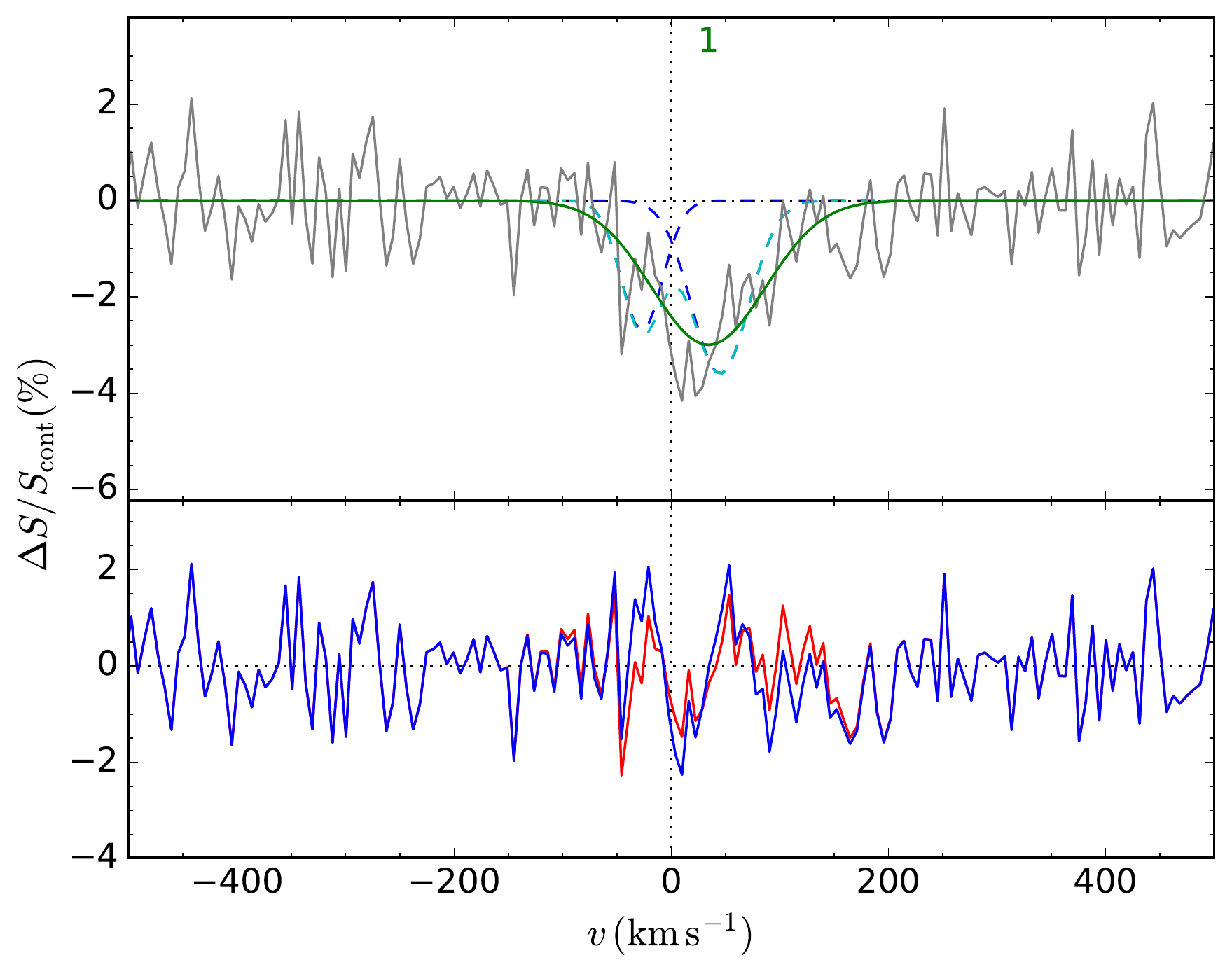}
\caption{The H{\sc i} absorption detection made towards PKS\,0500+019, centred here at $z$~=~0.58457 as in \citet{Carilli1998}. We do not recover the asymmetric absorption profile seen in \citet{Carilli1998} (blue dashed components, top panel). We instead find a single Gaussian component (green) to be the best fit to our observation. In the bottom panel we give the residuals of the data after subtraction of our single Gaussian component fit (red) and the two component fit from \citet{Carilli1998} (blue).}
\label{fig:PKS0500_fit}
\end{figure}

\begin{table*}
\normalsize
 \centering
  \caption{The inferred values for parameters from fitting a multiple Gaussian spectral-line model for the \mbox{H{\sc i}} absorption towards PKS\,1829$-$718, and the single Gaussian model fitted to PKS\,0500+019. $z_{\mathrm{centre}}$ is the redshift of the centre of the absorption component assuming a heliocentric standard-of-rest (HSR) frame; $\Delta{v}_{\mathrm{FWHM},i}$ is the FWHM; ($\Delta{S}/S_{\rm cont}$)$_{\rm peak}$ is the peak continuum depth as a fraction of the continuum flux density; and $R$ is the natural logarithm of the ratio of probability for this model versus the no spectral-line model (the Bayes odds ratio). This ratio was used to determine the number of components which best fitted the H{\sc i} absorption. Features are in order of decreasing redshift; that is, from right to left of Fig.~\ref{fig:PKS1829_fit} and \ref{fig:PKS0500_fit}. The parameters quoted for PKS\,1740$-$517 come from \citet{Allison2015}.}
  \begin{tabular}{lccccccc}
  \hline
Source name &$z_{\mathrm{centre}}$ & $\Delta{v}_{\mathrm{FWHM},i}$ & 1$\sigma$ error & ($\Delta{S}/S_{\rm cont}$)$_{\rm peak}$ & 1$\sigma$ error & $R$ \\
& & (km\,s$^{-1}$) & (km\,s$^{-1}$)& (\%) & (\%)& & \\
 \hline
 PKS\,0500+019 & 0.5848 & 126 & $^{+36}_{-31}$ & 2.8 & $^{+0.8}_{-0.8}$ & 2.9$\pm$0.1\\
 \hline
 PKS\,1740--517 & 0.4418 & 54 & $^{+5}_{-5}$ & 0.9 & $^{+0.1}_{-0.1}$ & 2868.19$\pm$0.3\\
 & 0.4413 & 5 & $^{+1}_{-1}$ & 20.4 & $^{+0.5}_{-0.6}$\\
 & 0.4412 & 8 & $^{+1}_{-1}$ & 4.1 & $^{+0.3}_{-0.3}$\\
 & 0.4406 & 338 & $^{+73}_{-64}$ & 0.2 & $^{+0.1}_{-0.1}$\\
 \hline
PKS\,1829--718 & 0.5364 & 46 & $^{+1}_{-1}$ & 24.6 & $^{+0.4}_{-0.9}$ & 5745.0$\pm$0.3\\
 & 0.5361 & 26 & $^{+2}_{-2}$ & 16.8 & $^{+1.0}_{-2.0}$\\
 & 0.5360 & 70 & $^{+3}_{-4}$ & 8.5 & $^{+0.8}_{-1.6}$\\
 & 0.5350 & 22 & $^{+2}_{-2}$ & 4.3 & $^{+0.4}_{-0.7}$\\
 & 0.5347 & 89 & $^{+17}_{-16}$ & 1.1 & $^{+0.2}_{-0.3}$\\
\hline
\end{tabular}
\label{tab:analysis_det}
\end{table*}

Fig.~\ref{fig:PKS1829_fit} is centred on the optical spectroscopic redshift we obtained ($z$~=~0.53467; Section~\ref{sec:opticalspect}). We concluded that the H{\sc i} absorption seen towards PKS\,1829$-$718 is intrinsic to the host galaxy of the radio AGN, and that the associated gas and dust is the source of quasar reddening. The systemic redshift aligns with Gaussian component 5 (Table~\ref{tab:analysis_det}). The remaining absorption features are redshifted up to $\sim$350~km\,s$^{-1}$; that is, moving away from the observer towards the nucleus. These redshifted components (1 to 4) may be gas clouds in-falling toward the AGN, as seen in \cite{VanGorkom1989}. We note that PKS\,1829$-$718 was identified as a GPS source in \cite{Callingham2017} with a peak frequency near 1~GHz, which typically indicates a linear extent of $<$1~kpc. 

We compare with two other examples of H{\sc i} absorption detections found in the literature found to have deep, redshifted absorption components relative to shallow features by a few hundred~km\,s$^{-1}$. It is noted that these absorption profiles are uncommon, hence the limited comparisons we can make. \cite{Srianand2015} observed two redshifted ($\sim$100-200\,km\,s$^{-1}$) H{\sc i} absorption features for radio source J094221.98+062335.2, relative to a deep absorption feature with peak optical depth $\sim$50\%. The two redshifted absorption features were far shallower with peak optical depths of a few percent. \cite{Srianand2015} suggested that as these shallow features were narrow (FWHM of $\sim$10\,km\,s$^{-1}$), the in-falling systems were clumpy gas clouds, rather than a smooth in-falling distribution, for which one would expect to detect broad absorption lines. The radio source in \cite{Srianand2015} is compact, and resolved into two symmetric cores separated by 89~mas through VLBA observations. The radio source is associated with a galaxy merger event. The two merging galaxies are separated by 4.8~kpc, and the H{\sc i} absorption could not be yet verified to be attributed to one galaxy or both. As all components for PKS\,1829$-$718 are narrow, with FWHMs of $<$100\,km\,s$^{-1}$, the redshifted gas seen may also be clumpy, and the obscuration of the galaxy attributed to a possible merger event; however, further higher-resolution observation will be required.

The second example of H{\sc i} absorption we compare with is that of NGC\,315 studied in \cite{Morganti2002}. Two absorption features had been detected: one shallow (optical depth of 1\%) at the systemic velocity with a FWHM of $\sim$80\,km\,s$^{-1}$, and a deep (optical depth of 23\%) and narrow feature redshifted by $\sim$500\,km\,s$^{-1}$. These properties are strikingly similar to components 5 and 1 for PKS\,1829$-$718 respectively. \cite{Morganti2002} postulated though the use of VLBA (Very Large Baseline Array) observations that the narrow deep feature is due to tidal debris. The shallow broader feature was suggested to be situated relatively close to the nucleus and distributed in a disk or torus. These interpretations could also hold for our observed absorption features. 

However, given that PKS\,1829$-$718 was unresolved for ASKAP-BETA and in the literature \cite[e.g. in AT20G;][]{Murphy2010}, we require higher angular resolution spectral line follow-up to confirm these possible explanations, and so this interpretation is merely a possibility. Another possible explanation for the asymmetric absorption features is that the galactic disk of PKS\,1829$-$718 is also asymmetric to our line of sight. We are unable to verify or rule out this from Fig.~\ref{fig:opticalimage} given the faintness of the source.

The centre of the radio position in the literature (NED) is offset from the optical by $\sim$1~arcsec. At this redshift this corresponds to a physical separation of $\sim$6~kpc. However, we stress that as PKS\,1829$-$718 is unresolved in radio observations here and in the literature (Fig.~\ref{fig:opticalimage}), we require higher angular resolution follow-up to better address both the interpretation of the H{\sc i} absorption and to verify any spatial offset. Comparing to the previously discussed examples, an offset of 0.2 arcseconds, or about $\sim$0.5~kpc, between the centre of radio emission seen in VLBI observations and optical SDSS position, was found for the source discussed in \citep{Srianand2015}. No offset is seen for the radio continuum and optical emission in the H{\sc i} absorption detection presented by \cite{Morganti2002}.

The optical spectrum has strong emission lines, indicative of a high excitation radio galaxy and a Type 2 AGN. Spectral energy distribution data in the literature for this source (e.g. NED) is limited to 22~GHz and below. We note that the source has a steep radio to mid-infrared spectral index ($\alpha^{\rm{rad}}_{\rm{MIR}}$~$<$~--0.8) when considering the faint WISE magnitudes for this source. However, there is potential confusion with a nearby source with WISE, making the WISE measurements, which includes upper limits in the W3 and W4 bands, unreliable. Nonetheless, the steep spectral index and lack of bright emission at higher frequencies is evident of strong reddening that may be occurring within the infrared regime as well as in the optical. 

We obtain an $E(B-V)$ value of 0.445~mag through template fitting from the Medium resolution INT Library of Empirical Spectra (MILES) stellar spectral library \citep{Vazdekis2010} to the optical spectrum we obtained, between the range of 4660--9396~{\AA}, via the Penalized Pixel-Fitting software package \cite[pPXF;][]{Cappellari2017}. For the reddening calculation we used the default option for pPXF, i.e. the extinction curve provided by \cite{Calzetti2000}. We note that this value is similar to $E(B-V)$ measurements (0.0--0.6~mag) seen for the radio-loud \cite[$>$500~mJy in the Faint Images of the Radio Sky at Twenty Centimeters - FIRST - survey;][]{White1997} red quasars from the FIRST--2MASS Red Quasar sample \cite[see fig. 11 and section 4.3 of][]{Glikman2012}. As proposed in \cite{Glikman2012}, this suggests that the optical faintness of the source associated with PKS\,1829$-$718 is due to dust extinction, rather than reddening attributed to synchrotron emission.

\subsubsection{PKS 1740$-$517}\label{sec:redest}

The H{\sc i} absorption detected toward PKS\,1740$-$517 was first reported and analysed by \cite{Allison2015}, and confirmed to be associated. This is a GPS source peaking around 1~GHz (middle panel of Fig.~\ref{fig:sed}). 

Like PKS\,1829$-$718, PKS\,1740$-$517 lacked spectroscopic redshift information prior to the detection of H{\sc i} absorption. A photometric redshift estimate of $z$~=~0.63 was obtained from the R-band magnitude of 20.8~mag by \cite{Burgess2006}, but the radio source redshift was confirmed to be at $z$~=~0.44 following ASKAP-BETA observations and optical follow-up \citep{Allison2015}. The H{\sc i} absorption was found to be associated with the radio source. It was inferred that the expanding young ($t_{\mathrm{age}}$~$\sim$~2500\,yr) radio source is cocooned within a dense medium, and may
be driving circumnuclear neutral gas in an outflow \cite[see also][]{Allison2019}.

\subsubsection{PKS 0500+019}

This well-studied variable radio source is know to have H{\sc i} absorption, which was originally detected by \cite{Carilli1998}. Significant steepening observed in the optical SED is consistent with very strong reddening \citep{Stickel1996}. It was also identified as a GPS source peaking at 1~GHz by \cite{Callingham2017} (left panel of Fig.~\ref{fig:sed}).

Prior to this study, while emission features ([O\,{\sc ii}] and [O\,{\sc iii}]) indicated a redshift of $z$~=~0.58, an unidentified emission line at higher redshift was also found toward this radio source \citep{Stickel1996}. However, this feature was stated by \cite{Kollgaard1995} to be possibly spurious, and the conclusion that the H{\sc i} absorption system and radio source are indeed associated, rather than PKS\,0500+019 being a background quasar, has since been supported \cite[e.g.][]{Stanghellini1997,deVries1998}.

We note that our 5-sigma optical depth sensitivity from our sole observation with ASKAP-12 is 3.7\%, which is greater than the optical depth of the shallower component detected in \cite{Carilli1998} ($\sim$2.7\%). While we still redetect the H{\sc i} absorption, we can only model our detection with a single Gaussian component (Fig.~\ref{fig:PKS0500_fit} and Table~\ref{tab:analysis_det}). The peak and integrated optical depths of the H{\sc i} absorption are $0.028 \pm 0.008$ and $\int \tau {\rm d}v$\,km~s$^{-1}$~=~$ 3.7 \pm 0.8$, respectively. Assuming a spin temperature of 100\,K, we obtain a column density of 6.7~$\times$~10$^{20}f$\,cm$^{-2}$. This is within errors of the value measured by \cite{Carilli1998} of 6.2~$\times$~10$^{20}$\,cm$^{-2}$ (Fig.~\ref{fig:column_flux}). Similar residuals are obtained when subtracting the Gaussian components given in \cite{Carilli1998} from our spectrum (Fig.~\ref{fig:PKS0500_fit}). 

\section{Discussion}\label{sec:5}

\subsection{HI absorption detection rate}

\subsubsection{Comparison with Carilli et al. (1998)}

Our H{\sc i} associated absorption detection rate of 12.5\%~$\pm$~6.8\% is significantly lower than the 80\%~$\pm$~17.9\% detection rate found in the five-target study by \cite{Carilli1998}. We note that this study had the same starting radio catalogue \citep{Kuhr1981} and R-band magnitude limit ($>$20~mag) as our optical sub-sample. Using our 12.5\% detection rate, the probability to replicate the 80\% detection rate of \cite{Carilli1998} with our sample is 0.1\%. This only slightly increases when assuming the associated detection rate for our optically faint subsample.

\begin{figure}
\includegraphics[width=1.0\linewidth]{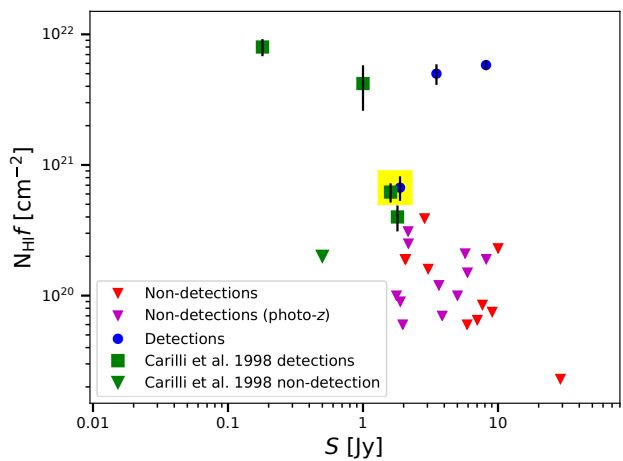}
\caption{The H{\sc i} column density (assuming a spin temperature of 100~K) versus the flux density of the source. Blue circles give our detections of H{\sc i} absorption. Red triangles give the upper limits of our non-detections with spectroscopic redshifts searched for associated absorption, while magenta triangles are sources in our sample with photometric redshift estimates.} Green squares are the \citet{Carilli1998} detections, and the green triangle the non-detection from the same study. Detection points within the yellow rectangle are the $N_{\rm{HI}}f$ measurements of PKS\,0500+019, observed both by us and \citet{Carilli1998}. The two measurements agree well. The non-detection upper limit in \citet{Carilli1998} is adjusted to give the 5-sigma upper limit, as with our upper limits, but retains the original assumed FWHM of 29~km\,s$^{-1}$. Our upper limits here assume a FWHM of 30~km\,s$^{-1}$.
\label{fig:column_flux}
\end{figure}

To investigate this discrepancy, in Fig.~\ref{fig:column_flux} we plot the H{\sc i} column density measurements multiplied by the covering factor (blue circles) and upper limits from the 5$\sigma$ optical depth sensitivity of the spectra (red and magenta triangles) with the median measured flux density (see Table~\ref{tab:derivedvalues}). These are compared with the H{\sc i} detections reported by \cite{Carilli1998} (green squares) which have the same spin temperature and equal covering factors for each source assumed for the comparison. We find that a few sources in our sample with no detections have a similar or lower H{\sc i} column density sensitivity to the detection with the lowest column density, and above the non-detection's upper limit, found by \cite{Carilli1998}. We were therefore potentially not sensitive enough in this early commissioning survey to detect H{\sc i} in some of these sources (down to $N_{\rm{HI}}$~$\sim$~2$\times$10$^{20}$~cm$^{-2}$). We note that we have adjusted the Carilli non-detection upper limits to a 5-sigma upper limit, to match our values. This slightly lower sensitivity may have only partially diminished our H{\sc i} absorption detection rate, and would not explain why the rate for \cite{Carilli1998} is so much higher than ours. 

As stated in Section~\ref{sec:3}, some observations suffered from correlator block failures, corrupting some $\sim$2\% of our total searched redshift space. This contributes to our lower H{\sc i} detection rate, albeit in a minor fashion. Lastly, it is possible that despite our photometric redshift estimates, some sources without spectroscopic redshift information were below the $z$~$<$~0.4 lower limit of our survey. 

All these factors must be considered in the comparison of the detection rates between our survey and that by \cite{Carilli1998}. Nonetheless, it does appear that the 80\% H{\sc i} absorption detection rate found by \cite{Carilli1998} towards reddened quasars is rather unusual even with all factors combined. It is possible that for the entire optically faint radio quasar population there is not as large a percentage with significant quantities of cold neutral gas as indicated by the \cite{Carilli1998} study. Further analysis of the full H{\sc i} absorption surveys with e.g. ASKAP, MeerKAT and Apertif, specifically towards obscured radio AGN, is necessary to further investigate this.

We find similar optical-to-NIR colours for the \cite{Carilli1998} H{\sc i} absorption detections to our detections. In the $R-W1$ colour space (Fig.~\ref{fig:iroptcolour}), \cite{Carilli1998} sources spans the range of 4.3--7.0~mag, with three of their detections having 6~$<$~$R-W1$~$<$~7~mag. Our three detections meanwhile span 6.2--7.2~mag. We however note the uncertain R-band magnitude available in the literature for PKS\,1829$-$718 \citep{Costa2001}, which could increase this range. If we consider both samples together that have a $R-W1$ magnitude between 6--7.5~mag, we have 5 H{\sc i} absorption detections out of 12, or a 42\ ~$\pm$~14\% detection rate. This suggests there is a higher presence of neutral atomic gas in obscured radio galaxies that have a steeper reddening profile in the optical-to-NIR regime.

This also could explain our lower H{\sc i} absorption detection rate, as many sources (including one \cite{Carilli1998} source) have lower $R-W1$ magnitudes. In our sample alone we find a 33\%~$\pm$~16\% (3/9) detection rate for sources with 6~$<$~$R-W1$~$<$~7.5~mag,
lower than the 75\%~$\pm$~22\% (3/4) rate found for the \cite{Carilli1998} sample in this same optical-to-NIR colour regime. Further investigation of the optical-to-NIR colour properties of H{\sc i} absorbers is required in future surveys to improve the small-number statistics available here.

\subsubsection{Non-detections in the WISE selected sub-sample}

No H{\sc i} absorption detection was made towards sources within the WISE selected sub-sample. Assuming a binomial probability equal to the 14.3\% ($\pm$~7.6\%) associated detection rate found for the optically selected sub-sample of 21 sources, we find a 21.3\% probability of zero associated detections to occur in the WISE-selected sub-sample of 10 sources at 0.4~$<$~$z$~$<$~1. This result is not statistically significant with this sample size. 


We nonetheless discuss a few possible factors that could have contributed to the lack of detections in this sub-sample. One is that by selecting only WISE sources with spectroscopic (8) or photometric (5) redshift estimates within the redshift range 0.4~$<$~$z$~$<$~1, we introduce a bias against either selecting quasars which are as obscured as the optically faint sub-sample, or higher redshift targets for intervening absorption. This may mean these sources in the WISE selected sub-sample have less obscuring dust, and cold gas associated with this dust, along the line of sight, although this would be surprising. Redder WISE AGN are believed to have higher star formation rates, ergo higher content which could obscure the AGN or be detected via H{\sc i} absorption. To account for this potential bias, we repeat the calculation for all 19 sources with WISE colour of W2--W3~$\geq$~3.5 (Table~\ref{tab:sourcelist} and Fig.~\ref{fig:wisecolours}), regardless of whether the source has no redshift information (1 source) or a redshift $z$~$>$~1 (5 sources). We find a 5.3\% probability of making no H{\sc i} absorption detections towards these 19 sources when assuming the 14.3\% detection rate, which has a corresponding p-value of 0.07; i.e. still a statistically insignificant result. This increases when using our total associated absorption detection rate of 12.5\%.

Another consideration is that the W2--W3 colour selection, and including sources with W1--W2~$\leq$~0.8 (albeit at lower priority), was ultimately not an effective method to target obscured quasars with cold neutral gas within their host galaxies. This is highlighted by combining our sample with the five \cite{Carilli1998} sources, where all six H{\sc i} absorption detections fall within the colour range of 2.5~$<$~$W2-W3$~$<$~3.3~mag - below our 3.5~mag cut for the WISE-selected sample. For all sources in the 2.5--3.3~mag range for $W2-W3$ colour we find a H{\sc i} absorption detection rate of 43\%~$\pm$~13\% (6/14). Larger sample sizes of an order of magnitude or greater, that will be made available through the upcoming full FLASH project, will be necessary to properly investigate the importance of the WISE mid-infrared colour properties of the H{\sc i} detections and non-detections. 

\subsubsection{Non-detections towards the $z$~$>$~1 sources}

No intervening H{\sc i} absorption detection was made toward the 10 sources with known spectroscopic redshifts above $z$~=~1 (bottom segment of Table~\ref{tab:sourcelist}, not including PKS\,0008-42). We investigate whether this result is unusual, or if we should have detected intervening H{\sc i} absorption for these sources. The number density of intervening DLA systems with a column density N$_{\rm{HI}}$~$>$~2$\times$10$^{20}$~cm$^{-2}$ \citep{Rao2017} is
\begin{equation}
N(z) = 0.027\,(1+z)^{1.682}.
\end{equation}
For a typical absorber redshift z~=~0.7, the number density computed from Equation~1 is 0.066. For a single sightline whose redshift search path is 0.6 (i.e. 0.4~$<$~z~$<$~1.0) this corresponds to a detection rate of 0.06~DLAs per sightline to a radio-bright AGN, assuming a spin temperature of 100~K. As the Poisson probability for no detections to be made is $\sim$55\% for 10 distinct sightlines to radio-bright AGN, this result is consistent with this probability. We note that using the equation given in \cite{Storrie-Lombardi2000} gives a 7\% probability for H{\sc i} absorption to be detected along one sightline, or $\sim$50\% for no H{\sc i} detections along 10 sightlines.

We next consider the full sample. We account for any spectroscopic redshift known in the calculation, and set that as the upper limit for the redshift to be probed for intervening absorption. For sources with no spectroscopic redshift known, we assume the photometric redshift estimate through the methods of \cite{Burgess2006b} and \cite{Glowacki2017b} (see Section~\ref{sec:redest}). In the case of PKS\,1213$-$17, we assume a host galaxy redshift of greater than one. Our total redshift pathlength searched for intervening H{\sc i} absorption, to a sensitivity of N$_{\rm{HI}}$~$\sim$~2$\times$10$^{20}$~cm$^{-2}$ and excluding regions affected by correlator block errors, is $\Delta\,z$~$\sim$~12. While this gives a number density of expected intervening DLAs of $\sim$2, the Poisson probability of no intervening detections in our whole sample is $\sim$13\%; again consistent with our result. We note that our two previously unknown H{\sc i} absorption detections were only verified to be associated from optical follow-up.

This result contrasts with those seen in the study of \cite{Gupta2009} and \cite{Gupta2012}. These two H{\sc i} absorption studies were solely toward strong Mg\,{\sc II} intervening sources. \cite{Gupta2009} found 9 out of 35 intervening H{\sc i} absorption detections (26\%) in the redshift range  1.10~$<$~$z$~$<$~1.45, while \cite{Gupta2012} presented 4 new detections in a sample of 17 Mg\,{\sc II} sources (24\%) in the redshift space 0~$<$~$z$~$<$~1.5. While only two of the detections made in \cite{Gupta2012} overlapped with our study's redshift range (0.4~$<$~$z$~$<$~1), they found that the 21-cm detection rate is fairly constant over 0.5~$<$~$z$~$<$~1.5. 

Most of our sources do not have known intervening Mg\,{\sc II} lines noted in the literature, nor were selected by this or related criteria. Only three of our sources at $z$~$>$~1 have an absorption line feature listed in the literature (NED) within our searchable redshift range of 0.4~$<$~$z$~$<$~1. Nonetheless, we consider the largest number density of 21-cm absorbers value quoted in \cite{Gupta2012} for 0.5~$<$~$z$~$<$~1 in table 5 ($n_{21}$~=~0.07\,$^{+0.10}_{-0.06}$ when combining their study with previous literature, for an integrated optical depth limit of 10\,km\,s${-1}$). This number density increases to 0.084 when adjusting to our larger redshift range (0.4~$<$~$z$~$<$~1), and assuming $n_{21}$ remains constant to this lower redshift. For our total redshift pathlength searched for H{\sc i} absorption, we find a Poisson probability of $\sim$37\% to make zero intervening H{\sc i} absorption detections. We highlight that these studies were more sensitive than ours; for example, the \cite{Gupta2009} study had H{\sc i} column densities ranging from 0.03--1.61~$\times\,$10$^{20}$~cm$^{-2}$, all lower than our detections by at least a factor of 3, and furthermore below many upper-limits of our non-detections (see Fig.~{\ref{fig:column_flux}} and Table~\ref{tab:derivedvalues}).

Out of 11 sources with $z$~$>$~1 in our study, 4 had upper limits N$_{\rm{HI}}$~$>$~2$\times$10$^{20}$~cm$^{-2}$ (assuming a spin temperature of 100~K), greater than the column density limit assumed in the number density found by \cite{Rao2017}. The number of sources with higher upper limits than 2$\times$10$^{20}$~cm$^{-2}$ increases to 9 sources if we assume a greater FWHM of 100\,km\,s$^{-1}$ (compared to 30\,km\,s$^{-1}$).  We note that this higher FWHM value is more commonly seen for associated H{\sc i} absorption than intervening cases \citep{Curran2016}. As higher column density DLAs are rarer, we have a lower effective detection rate (e.g. $\sim$5\% for N$_{\rm{HI}}$~$>$~1$\times$10$^{21}$~cm$^{-2}$, from integration of equation~6 in \cite{Noterdaeme2009}).  We also assumed a low spin temperature of 100~K; DLAs along the line-of-sight may instead have a higher spin temperature. These issues when considered together lower the probability to detect intervening H{\sc i} absorption. Nonetheless, we obtained upper limits on the H{\sc i} column density of these galaxies searched purely for intervening gas. Further H{\sc i} observations with ASKAP will aid our understanding of dusty intervening galaxies, and the distribution of cold neutral gas in such galaxies. 

We note that all three detections in this work show shallow ($<$5\% in optical depth) absorption features. Even when ignoring the presence of deeper absorption features (e.g. those seen for PKS\,1740$-$517 and PKS\,1829$-$718) and only considering the column densities of these shallow features, the above statistical analysis does not change significantly. We also note that the linefinder tool used in the analysis \citep{Allison2012b} is able to detect these broad shallow features, as shown in previous work. In \cite{Allison2012} a tentative broad, shallow feature was detected solely through the linefinder, and this was verified in follow-up observations \citep{Allison2013}. The linefinder tool will continue to support future ASKAP spectral line work.

%

\subsection{H{\sc i} detections in GPS sources}

All three radio sources in which we detect H{\sc i} are identified as GPS sources in \cite{Callingham2017} peaking around 1~GHz (Fig.~\ref{fig:sed}). As GPS radio sources are believed to be young or re-triggered radio jets, and PKS\,1740$-$517 has an estimated radio age of 2,500 years \citep{Allison2015}, we find support for the galaxy evolution model in \cite{Hopkins2006} which postulates that such obscured galaxies represent a short transitionary stage in galaxy evolution during which the AGN has only recently switched on \citep{Glikman2012}. That is, we detect H{\sc i} associated absorption towards obscured radio quasars likely to be young, frustrated (confined by a dense medium), or re-triggered AGN. 

Of our 12 GPS sources in the sample, four are at a redshift of $z$~$>$~1, although with PKS\,0008$-$42 we were able to still search for associated absorption by observing at a lower frequency (Section~\ref{sec:3}). We have no spectroscopic redshift information for PKS\,2333$-$528, but photometric redshift estimation suggests it lies within the redshift range 0.4~$<$~$z$~$<$~1. If included, we have a 33\% H{\sc i} associated absorption detection rate towards GPS sources, and 38\% if excluded ($\pm$16\%). Including the steep spectrum sources not known to be at $z$~$>$~1 and excluding PKS\,0008$-$42 (which we searched for associated H{\sc i} absorption), we found a 14\% ($\pm$7\%) associated detection rate, although this assumes the 12 other radio quasars in our sample without spectroscopic redshifts lie between 0.4~$<$~$z$~$<$~1. If we exclude sources with only photometric redshifts, the detection rate becomes 27\% ($\pm$13\%). 

Our detection rate range agrees with that of other studies towards compact sources, including GPS and CSS selected samples \cite[e.g. with detection rates of $\sim$10--40\% in][]{Vermeulen2003, Gupta2006, Chandola2011, Glowacki2017} and slightly higher than for compact flat spectrum sources \cite[e.g. detection rate of 17\% in the Caltech--Jodrell flat-spectrum sample;][]{Aditya2018}. Our detections made towards the GPS sources also support the findings of \cite{Maccagni2017}, who find a higher H{\sc i} detection rate towards compact radio galaxies (32\% versus 16\% for the extended sources within their sample). Our GPS sources could be oriented such that the line of sight goes through the dusty torus, hence increasing our chance of detecting H{\sc i} absorption. A consideration of the radio SED \citep{Callingham2017} and linear extent, as well as available optical and mid-infrared information, is important for future searches of H{\sc i} absorption in AGN obscured by material in the host galaxy to better classify the galaxy, investigate the evolution of radio sources, and identify redshifts. 

\cite{Gereb2015} and \cite{Glowacki2017} found shallow (optical depths $<$ 5\%), broad (FWHM upward of 100~km\,s$^{-1}$) and disturbed (asymmetric) H{\sc i} absorption associated with compact radio galaxies at lower redshift (0.02~$<$~$z$~$<$~0.23 and 0.04~$<$~$z$~$<$~0.1 respectively). Likewise, we find shallow absorption features in all three of our detections, and two with asymmetric H{\sc i} absorption features for PKS\,1740$-$517 and PKS\,1829$-$718. This supports the hypothesis of \cite{Gereb2015} that the cold neutral medium in compact radio galaxies is kinematically disturbed, as postulated for PKS\,1740$-$517 \citep{Allison2015}.

Only two of the four detections made by \cite{Carilli1998} towards reddened quasars had absorption features with optical depths below 5\%. We therefore see a higher percentage of shallow absorption features in our H{\sc i} absorption detections. Three of the four H{\sc i} absorption detections in \cite{Carilli1998} are asymmetric. Blue-shifted and asymmetric H{\sc i} absorption features were also seen in the single detection reported in \cite{Ishwara-Chandra2003}, who proposed that the neutral gas has been disturbed by a radio jet. These combined results suggest that obscured/potentially young radio AGN, through feedback processes such as radio jets, may have a disruptive effect on the neutral star forming reservoir up to at least redshift $z$~$\sim$~1.

\section{Conclusion}

We conducted an early science H{\sc i} absorption commissioning survey with the ASKAP-BETA prototype array toward obscured quasars selected by either their optical obscuration or those with redder mid-infrared colour. We searched for H{\sc i} absorption that was associated with the host galaxy of the AGN to investigate the impact of AGN feedback on the cold neutral gas. For some sources we could only search for intervening H{\sc i} absorption (e.g. gas intervening DLAs along the line of sight), which is an alternate cause for obscuration of the background radio quasar. Lastly, for some sources observed we had no prior redshift information. 

We report three H{\sc i} absorption detections in our sample, one toward PKS\,1829$-$718 that was previously unknown, another toward PKS\,1740$-$517 \citep{Allison2015}, and the third a re-detection of one toward PKS\,0500+019 \citep{Carilli1998}. All three detections are made towards sources selected by optical faintness, and we see no H{\sc i} absorption towards sources selected by their WISE mid-infrared colours. However, the lack of detections in the WISE selected sub-sample is not yet statistically significant, and further analysis during the full FLASH project is required. 

Through optical spectroscopic follow-up and information within the literature for PKS\,1829$-$718, PKS\,1740$-$517 and PKS\,0500+019, we find that the detected H{\sc i} absorption is associated with the host galaxy of the AGN in each case. These three AGN are obscured by material in the host galaxy, e.g. a dusty torus formed following a galaxy merger event, or perhaps in a larger disk on the scale of $\sim$100~pc. 

We do not detect any H{\sc i} absorption towards our higher redshift targets ($z$~$>$~1). This is not necessarily unexpected given the predicted number of DLAs along the line of sight for our redshift range of 0.4~$<$~$z$~$<$~1. One factor which negatively contributed to this include the occasional correlator block error resulting in corrupted segments of the spectrum. Another is that some targets had limited integration time resulting in higher optical depth limits, and corresponding higher H{\sc i} column density upper limits (Table~\ref{tab:obslist}). This lowered our sensitivity to detecting H{\sc i} gas in absorption.

All detections are made towards GPS sources, and we find a 33\% H{\sc i} associated absorption detection rate towards the GPS sources in our sample.  These GPS sources are compact in linear extent and believed to be young or re-triggered in their AGN activity. We also find shallow and asymmetric H{\sc i} absorption features in our survey, as seen in \cite{Carilli1998} and \cite{Ishwara-Chandra2003} towards red quasars, and low redshift studies towards young radio AGN.

The full FLASH project, alongside other future all-sky H{\sc i} absorption surveys (MALS and SHARP), will better increase our H{\sc i} statistics. This in turn will improve our understanding of the impact of the radio AGN on its cold gas, and its evolution with redshift, during this crucial transitional stage in galaxy evolution. Intervening H{\sc i} absorption studies alongside the other ASKAP survey science projects will increase our statistics of the obscured radio AGN population missed in optically selected samples, and understanding of the distribution of cold neutral gas with redshift for such sources.

\section*{Acknowledgements}

We wish to thank the anonymous referee for their useful feedback, which helped improve this paper. We thank  Nathan Deg, Samuel Hinton and Nicholas Scott for useful discussions. We also thank Robert Beswick, Jeremy Darling, and John Dickey for their comments on a thesis chapter version of this work. Parts of this research were conducted by the Australian Research Council Centre of Excellence for All-sky Astrophysics (CAASTRO) and the Australian Research Council Centre of Excellence for All Sky Astrophysics in 3 Dimensions (ASTRO 3D), through project numbers CE110001020 and CE170100013 respectively. We gratefully acknowledge and thank the ASKAP commissioning team for the use of BETA when time was available, thus enabling the studies described in this paper. The Australian Square Kilometre Array Pathfinder (ASKAP) is part of the Australia Telescope National Facility which is managed by CSIRO. Operation of ASKAP is funded by the Australian Government with support from the National Collaborative Research Infrastructure Strategy. ASKAP uses the resources of the Pawsey Supercomputing Centre. Establishment of ASKAP, the Murchison Radio-astronomy Observatory and the Pawsey Supercomputing Centre are initiatives of the Australian Government, with support from the Government of Western Australia and the Science and Industry Endowment Fund. We acknowledge the Wajarri Yamatji people as the traditional owners of the Observatory site.

This research has made use of the NASA/IPAC Extragalactic Database (NED) and Infrared Science Archive, which are operated by the Jet Propulsion Laboratory, California Institute of Technology, under contract with the National Aeronautics and Space Administration. This publication makes use of data products from the Wide-field Infrared Survey Explorer \citep{Wright2010}, which is a joint project of the University of California, Los Angeles. We also acknowledge use of APLpy, an open-source plotting package for Python hosted at http://aplpy.github.com \citep{Robitaille2012}, and Astropy, a community-developed core Python package for Astronomy (Astropy Collaboration, 2013). This research has made use of SAOImage DS9, developed by Smithsonian Astrophysical Observatory. This work includes observations obtained at the Gemini Observatory, which is operated by the Association of Universities for Research in Astronomy, Inc., under a cooperative agreement with the NSF on behalf of the Gemini partnership: the National Science Foundation (United States), the National Research Council (Canada), CONICYT (Chile), Ministerio de Ciencia, Tecnolog\'{i}a e Innovaci\'{o}n Productiva (Argentina), and Minist\'{e}rio da Ci\^{e}ncia, Tecnologia e Inova\c{c}\~{a}o (Brazil).

\footnotesize{
  \bibliographystyle{mn2e}
  \bibliography{bibliography}

\begin{thebibliography}{128}
\expandafter\ifx\csname natexlab\endcsname\relax\def\natexlab#1{#1}\fi

\bibitem[{{Aditya} \& {Kanekar}(2018)}]{Aditya2018}
{Aditya} J.~N.~H.~S., {Kanekar} N., 2018, MNRAS, 481, 1578

\bibitem[{Allison {et~al}\mbox{.}(2012{\natexlab{a}})Allison, Curran, Emonts,
  Ger\'{e}b, Mahony, Reeves, Sadler, Tanna, Whiting, \& Zwaan}]{Allison2012}
Allison J.~R. {et~al.}, 2012{\natexlab{a}}, MNRAS, 423, 2601

\bibitem[{Allison {et~al}\mbox{.}(2013)Allison, Curran, Sadler, \&
  Reeves}]{Allison2013}
Allison J.~R., Curran S.~J., Sadler E.~M., Reeves S.~N., 2013, MNRAS, 430, 157

\bibitem[{{Allison} {et~al}\mbox{.}(2019){Allison}, {Mahony}, {Moss}, {Sadler},
  {Whiting}, {Allison}, {Bland- Hawthorn}, {Curran}, {Emonts}, {Lagos},
  {Morganti}, {Tremblay}, {Zwaan}, {Anderson}, {Bunton}, \&
  {Voronkov}}]{Allison2019}
{Allison} J.~R. {et~al.}, 2019, MNRAS, 482, 2934

\bibitem[{Allison {et~al}\mbox{.}(2015)Allison, Sadler, Moss, Whiting,
  Hunstead, Pracy, Curran, Croom, Glowacki, Morganti, Shabala, Zwaan, Allen,
  Amy, Axtens, Ball, Bannister, Barker, Bell, Bock, Bolton, Bowen, Boyle,
  Braun, Broadhurst, Brodrick, Brothers, Brown, Bunton, Cantrall, Chapman,
  Cheng, Chippendale, Chung, Cooray, Cornwell, DeBoer, Diamond, Edwards, Ekers,
  Feain, Ferris, Forsyth, Gough, Grancea, Gupta, Guzman, Hampson, Harvey-Smith,
  Haskins, Hay, Hayman, Heywood, Hotan, Hoyle, Humphreys, Indermuehle, Jacka,
  Jackson, Jackson, Jeganathan, Johnston, Joseph, Kendall, Kesteven, Kiraly,
  Koribalski, Leach, Lenc, Lensson, Mackay, Macleod, Marquarding, Marvil,
  McClure-Griffiths, McConnell, Mirtschin, Norris, Neuhold, Ng, O'Sullivan,
  Pathikulangara, Pearce, Phillips, Popping, Qiao, Reynolds, Roberts, Sault,
  Schinckel, Serra, Shaw, Shields, Shimwell, Storey, Sweetnam, Troup, Turner,
  Tuthill, Tzioumis, Voronkov, Westmeier, \& Wilson}]{Allison2015}
Allison J.~R. {et~al.}, 2015, MNRAS, 453, 1249

\bibitem[{Allison {et~al}\mbox{.}(2012{\natexlab{b}})Allison, Sadler, \&
  Whiting}]{Allison2012b}
Allison J. R.~A., Sadler E. M.~A., Whiting M. T.~C., 2012{\natexlab{b}}, PASA,
  29, 221

\bibitem[{{Assef} {et~al}\mbox{.}(2013){Assef}, {Stern}, {Kochanek}, {Blain},
  {Brodwin}, {Brown}, {Donoso}, {Eisenhardt}, {Jannuzi}, {Jarrett}, {Stanford},
  {Tsai}, {Wu}, \& {Yan}}]{Assef2013}
{Assef} R.~J. {et~al.}, 2013, ApJ, 772, 26

\bibitem[{{Best} {et~al}\mbox{.}(1999){Best}, {R{\"o}ttgering}, \&
  {Lehnert}}]{Best1999}
{Best} P.~N., {R{\"o}ttgering} H.~J.~A., {Lehnert} M.~D., 1999, MNRAS, 310, 223

\bibitem[{{Blecha} {et~al}\mbox{.}(2018){Blecha}, {Snyder}, {Satyapal}, \&
  {Ellison}}]{Blecha2018}
{Blecha} L., {Snyder} G.~F., {Satyapal} S., {Ellison} S.~L., 2018, MNRAS, 478,
  3056

\bibitem[{{Booth} \& {Jonas}(2012)}]{Booth2012}
{Booth} R.~S., {Jonas} J.~L., 2012, African Skies, 16, 101

\bibitem[{{Burgess} \& {Hunstead}(2006{\natexlab{a}})}]{Burgess2006}
{Burgess} A.~M., {Hunstead} R.~W., 2006{\natexlab{a}}, AJ, 131, 100

\bibitem[{{Burgess} \& {Hunstead}(2006{\natexlab{b}})}]{Burgess2006b}
{Burgess} A.~M., {Hunstead} R.~W., 2006{\natexlab{b}}, AJ, 131, 114

\bibitem[{{Callingham} {et~al}\mbox{.}(2017){Callingham}, {Ekers}, {Gaensler},
  {Line}, {Hurley-Walker}, {Sadler}, {Tingay}, {Hancock}, {Bell},
  {Dwarakanath}, {For}, {Franzen}, {Hindson}, {Johnston-Hollitt},
  {Kapi{\'n}ska}, {Lenc}, {McKinley}, {Morgan}, {Offringa}, {Procopio},
  {Staveley-Smith}, {Wayth}, {Wu}, \& {Zheng}}]{Callingham2017}
{Callingham} J.~R. {et~al.}, 2017, ApJ, 836, 174

\bibitem[{{Callingham} {et~al}\mbox{.}(2015){Callingham}, {Gaensler}, {Ekers},
  {Tingay}, {Wayth}, {Morgan}, {Bernardi}, {Bell}, {Bhat}, {Bowman}, {Briggs},
  {Cappallo}, {Deshpande}, {Ewall-Wice}, {Feng}, {Greenhill}, {Hazelton},
  {Hindson}, {Hurley-Walker}, {Jacobs}, {Johnston-Hollitt}, {Kaplan},
  {Kudrayvtseva}, {Lenc}, {Lonsdale}, {McKinley}, {McWhirter}, {Mitchell},
  {Morales}, {Morgan}, {Oberoi}, {Offringa}, {Ord}, {Pindor}, {Prabu},
  {Procopio}, {Riding}, {Srivani}, {Subrahmanyan}, {Udaya Shankar}, {Webster},
  {Williams}, \& {Williams}}]{Callingham2015}
{Callingham} J.~R. {et~al.}, 2015, ApJ, 809, 168

\bibitem[{{Calzetti} {et~al}\mbox{.}(2000){Calzetti}, {Armus}, {Bohlin},
  {Kinney}, {Koornneef}, \& {Storchi-Bergmann}}]{Calzetti2000}
{Calzetti} D., {Armus} L., {Bohlin} R.~C., {Kinney} A.~L., {Koornneef} J.,
  {Storchi-Bergmann} T., 2000, ApJ, 533, 682

\bibitem[{{Cappellari}(2017)}]{Cappellari2017}
{Cappellari} M., 2017, \mnras, 466, 798

\bibitem[{Carilli {et~al}\mbox{.}(1998)Carilli, Menten, Reid, Rupen, \&
  Yun}]{Carilli1998}
Carilli C., Menten K., Reid M.~J., Rupen M., Yun M., 1998, ApJ, 494, 175

\bibitem[{{Carilli} {et~al}\mbox{.}(1999){Carilli}, {Menten}, \&
  {Moore}}]{Carilli1999}
{Carilli} C.~L., {Menten} K.~M., {Moore} C.~P., 1999, in Astronomical Society
  of the Pacific Conference Series, Vol. 156, Highly Redshifted Radio Lines,
  {Carilli} C.~L., {Radford} S.~J.~E., {Menten} K.~M., {Langston} G.~I., eds.,
  p. 171

\bibitem[{Catinella {et~al}\mbox{.}(2008)Catinella, Haynes, Giovanelli,
  Gardner, \& Connolly}]{Catinella2008}
Catinella B., Haynes M.~P., Giovanelli R., Gardner J.~P., Connolly A.~J., 2008,
  ApJ, 685, L13

\bibitem[{Chandola {et~al}\mbox{.}(2011)Chandola, Sirothia, \&
  Saikia}]{Chandola2011}
Chandola Y., Sirothia S.~K., Saikia D.~J., 2011, MNRAS, 418, 1787

\bibitem[{{Chengalur} \& {Kanekar}(2000)}]{Chengalur2000}
{Chengalur} J.~N., {Kanekar} N., 2000, MNRAS, 318, 303

\bibitem[{{Chippendale} {et~al}\mbox{.}(2015){Chippendale}, {Brown},
  {Beresford}, {Hampson}, {Macleod}, {Shaw}, {Brothers}, {Cantrall}, {Forsyth},
  {Hay}, \& {Leach}}]{Chippendale2015}
{Chippendale} A.~P. {et~al.}, 2015, in 2015 International Conference on
  Electromagnetics in Advanced Applications (ICEAA), p. 541-544, pp. 541--544

\bibitem[{{Cluver} {et~al}\mbox{.}(2014){Cluver}, {Jarrett}, {Hopkins},
  {Driver}, {Liske}, {Gunawardhana}, {Taylor}, {Robotham}, {Alpaslan},
  {Baldry}, {Brown}, {Peacock}, {Popescu}, {Tuffs}, {Bauer}, {Bland-Hawthorn},
  {Colless}, {Holwerda}, {Lara-L{\'o}pez}, {Leschinski},
  {L{\'o}pez-S{\'a}nchez}, {Norberg}, {Owers}, {Wang}, \&
  {Wilkins}}]{Cluver2014}
{Cluver} M.~E. {et~al.}, 2014, ApJ, 782, 90

\bibitem[{{Collier} {et~al}\mbox{.}(2014){Collier}, {Banfield}, {Norris},
  {Schnitzeler}, {Kimball}, {Filipovi{\'c}}, {Jarrett}, {Lonsdale}, \&
  {Tothill}}]{Collier2014}
{Collier} J.~D. {et~al.}, 2014, MNRAS, 439, 545

\bibitem[{Condon {et~al}\mbox{.}(1998)Condon, Cotton, Greisen, Yin, Perley,
  Taylor, \& Broderick}]{Condon1998}
Condon J.~J., Cotton W.~D., Greisen E.~W., Yin Q.~F., Perley R.~A., Taylor
  G.~B., Broderick J.~J., 1998, AJ, 115, 1693

\bibitem[{{Costa}(2001)}]{Costa2001}
{Costa} E., 2001, A\&A, 367, 719

\bibitem[{{Curran}(2018)}]{Curran2018}
{Curran} S.~J., 2018, arXiv:1807.08860

\bibitem[{{Curran} {et~al}\mbox{.}(2016){Curran}, {Duchesne}, {Divoli}, \&
  {Allison}}]{Curran2016}
{Curran} S.~J., {Duchesne} S.~W., {Divoli} A., {Allison} J.~R., 2016, MNRAS,
  462, 4197

\bibitem[{Curran {et~al}\mbox{.}(2006)Curran, Whiting, Murphy, Webb, Longmore,
  Pihlstr\"{o}m, Athreya, \& Blake}]{Curran2006}
Curran S.~J., Whiting M.~T., Murphy M.~T., Webb J.~K., Longmore S.~N.,
  Pihlstr\"{o}m Y.~M., Athreya R., Blake C., 2006, MNRAS, 371, 431

\bibitem[{{de Vries} {et~al}\mbox{.}(2007){de Vries}, {Snellen}, {Schilizzi},
  {Lehnert}, \& {Bremer}}]{deVries2007}
{de Vries} N., {Snellen} I.~A.~G., {Schilizzi} R.~T., {Lehnert} M.~D., {Bremer}
  M.~N., 2007, A\&A, 464, 879

\bibitem[{{de Vries} {et~al}\mbox{.}(1995){de Vries}, {Barthel}, \&
  {Hes}}]{deVries1995}
{de Vries} W.~H., {Barthel} P.~D., {Hes} R., 1995, AAPS, 114, 259

\bibitem[{{de Vries} {et~al}\mbox{.}(1998){de Vries}, {O'Dea}, {Perlman},
  {Baum}, {Lehnert}, {Stocke}, {Rector}, \& {Elston}}]{deVries1998}
{de Vries} W.~H., {O'Dea} C.~P., {Perlman} E., {Baum} S.~A., {Lehnert} M.~D.,
  {Stocke} J., {Rector} T., {Elston} R., 1998, ApJ, 503, 138

\bibitem[{Deboer {et~al}\mbox{.}(2009)Deboer, Gough, Bunton, Cornwell,
  Beresford, Johnston, Feain, Schinckel, Jackson, Kesteven, Chippendale,
  Hampson, Sullivan, Hay, Jacka, Sweetnam, Storey, Ball, \& Boyle}]{Deboer2009}
Deboer B. D.~R. {et~al.}, 2009, Proc. IEEE, 97, 1507

\bibitem[{{di Serego-Alighieri} {et~al}\mbox{.}(1994){di Serego-Alighieri},
  {Danziger}, {Morganti}, \& {Tadhunter}}]{diSerego-Alighieri1994}
{di Serego-Alighieri} S., {Danziger} I.~J., {Morganti} R., {Tadhunter} C.~N.,
  1994, MNRAS, 269, 998

\bibitem[{{Dutta} {et~al}\mbox{.}(2018){Dutta}, {Srianand}, \&
  {Gupta}}]{Dutta2018}
{Dutta} R., {Srianand} R., {Gupta} N., 2018, MNRAS, 480, 947

\bibitem[{{Ellison} {et~al}\mbox{.}(2018){Ellison}, {Catinella}, \&
  {Cortese}}]{Ellison2018}
{Ellison} S.~L., {Catinella} B., {Cortese} L., 2018, MNRAS, 478, 3447

\bibitem[{{Ellison} {et~al}\mbox{.}(2004){Ellison}, {Churchill}, {Rix}, \&
  {Pettini}}]{Ellison2004}
{Ellison} S.~L., {Churchill} C.~W., {Rix} S.~A., {Pettini} M., 2004, ApJ, 615,
  118

\bibitem[{{Ellison} {et~al}\mbox{.}(2012){Ellison}, {Kanekar}, {Prochaska},
  {Momjian}, \& {Worseck}}]{Ellison2012}
{Ellison} S.~L., {Kanekar} N., {Prochaska} J.~X., {Momjian} E., {Worseck} G.,
  2012, MNRAS, 424, 293

\bibitem[{{Ellison} {et~al}\mbox{.}(2011){Ellison}, {Nair}, {Patton},
  {Scudder}, {Mendel}, \& {Simard}}]{Ellison2011}
{Ellison} S.~L., {Nair} P., {Patton} D.~R., {Scudder} J.~M., {Mendel} J.~T.,
  {Simard} L., 2011, MNRAS, 416, 2182

\bibitem[{{Ellison} {et~al}\mbox{.}(2015){Ellison}, {Patton}, \&
  {Hickox}}]{Ellison2015}
{Ellison} S.~L., {Patton} D.~R., {Hickox} R.~C., 2015, MNRAS, 451, L35

\bibitem[{{Ellison} {et~al}\mbox{.}(2010){Ellison}, {Prochaska}, {Hennawi},
  {Lopez}, {Usher}, {Wolfe}, {Russell}, \& {Benn}}]{Ellison2010}
{Ellison} S.~L., {Prochaska} J.~X., {Hennawi} J., {Lopez} S., {Usher} C.,
  {Wolfe} A.~M., {Russell} D.~M., {Benn} C.~R., 2010, MNRAS, 406, 1435

\bibitem[{{Ellison} {et~al}\mbox{.}(2001){Ellison}, {Yan}, {Hook}, {Pettini},
  {Wall}, \& {Shaver}}]{Ellison2001}
{Ellison} S.~L., {Yan} L., {Hook} I.~M., {Pettini} M., {Wall} J.~V., {Shaver}
  P., 2001, A\&A, 379, 393

\bibitem[{{Fern{\'a}ndez} {et~al}\mbox{.}(2016){Fern{\'a}ndez}, {Gim}, {van
  Gorkom}, {Yun}, {Momjian}, {Popping}, {Chomiuk}, {Hess}, {Hunt}, {Kreckel},
  {Lucero}, {Maddox}, {Oosterloo}, {Pisano}, {Verheijen}, {Hales}, {Chung},
  {Dodson}, {Golap}, {Gross}, {Henning}, {Hibbard}, {Jaff{\'e}}, {Donovan
  Meyer}, {Meyer}, {Sanchez-Barrantes}, {Schiminovich}, {Wicenec}, {Wilcots},
  {Bershady}, {Scoville}, {Strader}, {Tremou}, {Salinas}, \&
  {Ch{\'a}vez}}]{Fernandez2016}
{Fern{\'a}ndez} X. {et~al.}, 2016, ApJL, 824, L1

\bibitem[{{Francis} {et~al}\mbox{.}(2000){Francis}, {Whiting}, \&
  {Webster}}]{Francis2000}
{Francis} P.~J., {Whiting} M.~T., {Webster} R.~L., 2000, PASA, 17, 56

\bibitem[{Freudling {et~al}\mbox{.}(2011)Freudling, Staveley-Smith, Catinella,
  Minchin, Calabretta, Momjian, Zwaan, Meyer, \& O'Neil}]{Freudling2011}
Freudling W. {et~al.}, 2011, ApJ, 727, 40

\bibitem[{{Ge} \& {Bechtold}(1997)}]{Ge1997}
{Ge} J., {Bechtold} J., 1997, ApJL, 477, L73

\bibitem[{{Ge} \& {Bechtold}(1999)}]{Ge1999}
{Ge} J., {Bechtold} J., 1999, in Astronomical Society of the Pacific Conference
  Series, Vol. 156, Highly Redshifted Radio Lines, {Carilli} C.~L., {Radford}
  S.~J.~E., {Menten} K.~M., {Langston} G.~I., eds., p. 121

\bibitem[{Ger\'{e}b {et~al}\mbox{.}(2015)Ger\'{e}b, Maccagni, Morganti, \&
  Oosterloo}]{Gereb2015}
Ger\'{e}b K., Maccagni F.~M., Morganti R., Oosterloo T.~A., 2015, A\&A, 575,
  A44

\bibitem[{{Giovanelli} {et~al}\mbox{.}(2005){Giovanelli}, {Haynes}, {Kent},
  {Perillat}, {Saintonge}, {Brosch}, {Catinella}, {Hoffman}, {Stierwalt},
  {Spekkens}, {Lerner}, {Masters}, {Momjian}, {Rosenberg}, {Springob},
  {Boselli}, {Charmandaris}, {Darling}, {Davies}, {Garcia Lambas}, {Gavazzi},
  {Giovanardi}, {Hardy}, {Hunt}, {Iovino}, {Karachentsev}, {Karachentseva},
  {Koopmann}, {Marinoni}, {Minchin}, {Muller}, {Putman}, {Pantoja}, {Salzer},
  {Scodeggio}, {Skillman}, {Solanes}, {Valotto}, {van Driel}, \& {van
  Zee}}]{Giovanelli2005}
{Giovanelli} R. {et~al.}, 2005, AJ, 130, 2598

\bibitem[{{Glikman} {et~al}\mbox{.}(2007){Glikman}, {Helfand}, {White},
  {Becker}, {Gregg}, \& {Lacy}}]{Glikman2007}
{Glikman} E., {Helfand} D.~J., {White} R.~L., {Becker} R.~H., {Gregg} M.~D.,
  {Lacy} M., 2007, ApJ, 667, 673

\bibitem[{{Glikman} {et~al}\mbox{.}(2012){Glikman}, {Urrutia}, {Lacy},
  {Djorgovski}, {Mahabal}, {Myers}, {Ross}, {Petitjean}, {Ge}, {Schneider}, \&
  {York}}]{Glikman2012}
{Glikman} E. {et~al.}, 2012, ApJ, 757, 51

\bibitem[{{Glowacki} {et~al}\mbox{.}(2017{\natexlab{a}}){Glowacki}, {Allison},
  {Sadler}, {Moss}, {Curran}, {Musaeva}, {Deng}, {Parry}, \&
  {Sligo}}]{Glowacki2017}
{Glowacki} M. {et~al.}, 2017{\natexlab{a}}, MNRAS, 467, 2766

\bibitem[{{Glowacki} {et~al}\mbox{.}(2017{\natexlab{b}}){Glowacki}, {Allison},
  {Sadler}, {Moss}, \& {Jarrett}}]{Glowacki2017b}
{Glowacki} M., {Allison} J.~R., {Sadler} E.~M., {Moss} V.~A., {Jarrett} T.~H.,
  2017{\natexlab{b}}, ArXiv:1709.08634

\bibitem[{{Gregg} {et~al}\mbox{.}(2002{\natexlab{a}}){Gregg}, {Lacy}, {White},
  {Glikman}, {Helfand}, {Becker}, \& {Brotherton}}]{Gregg2002}
{Gregg} M.~D., {Lacy} M., {White} R.~L., {Glikman} E., {Helfand} D., {Becker}
  R.~H., {Brotherton} M.~S., 2002{\natexlab{a}}, ApJ, 564, 133

\bibitem[{{Gregg} {et~al}\mbox{.}(2002{\natexlab{b}}){Gregg}, {Lacy}, {White},
  {Glikman}, {Helfand}, {Becker}, \& {Brotherton}}]{Gregg2001}
{Gregg} M.~D., {Lacy} M., {White} R.~L., {Glikman} E., {Helfand} D., {Becker}
  R.~H., {Brotherton} M.~S., 2002{\natexlab{b}}, \apj, 564, 133

\bibitem[{Gupta {et~al}\mbox{.}(2006)Gupta, Salter, Saikia, Ghosh, \&
  Jeyakumar}]{Gupta2006}
Gupta N., Salter C.~J., Saikia D.~J., Ghosh T., Jeyakumar S., 2006, MNRAS, 373,
  972

\bibitem[{{Gupta} {et~al}\mbox{.}(2017){Gupta}, {Srianand}, {Baan}, {Baker},
  {Beswick}, {Bhatnagar}, {Bhattacharya}, {Bosma}, {Carilli}, {Cluver},
  {Combes}, {Cress}, {Dutta}, {Fynbo}, {Heald}, {Hilton}, {Hussain}, {Jarvis},
  {Jozsa}, {Kamphuis}, {Kembhavi}, {Kerp}, {Kl{\"o}ckner}, {Krogager},
  {Kulkarni}, {Ledoux}, {Mahabal}, {Mauch}, {Moodley}, {Momjian}, {Morganti},
  {Noterdaeme}, {Oosterloo}, {Petitjean}, {Schr{\"o}der}, {Serra}, {Sievers},
  {Spekkens}, {V{\"a}is{\"a}nen}, {van der Hulst}, {Vivek}, {Wang}, {Wong}, \&
  {Zungu}}]{Gupta2017}
{Gupta} N. {et~al.}, 2017, ArXiv:1708.07371

\bibitem[{{Gupta} {et~al}\mbox{.}(2012){Gupta}, {Srianand}, {Petitjean},
  {Bergeron}, {Noterdaeme}, \& {Muzahid}}]{Gupta2012}
{Gupta} N., {Srianand} R., {Petitjean} P., {Bergeron} J., {Noterdaeme} P.,
  {Muzahid} S., 2012, A\&A, 544, A21

\bibitem[{{Gupta} {et~al}\mbox{.}(2009){Gupta}, {Srianand}, {Petitjean},
  {Noterdaeme}, \& {Saikia}}]{Gupta2009}
{Gupta} N., {Srianand} R., {Petitjean} P., {Noterdaeme} P., {Saikia} D.~J.,
  2009, MNRAS, 398, 201

\bibitem[{Hambly {et~al}\mbox{.}(2001)Hambly, MacGillivray, Read, Tritton,
  Thomson, Kelly, Morgan, Smith, Driver, Williamson, Parker, Hawkins, Williams,
  \& Lawrence}]{Hambly2001}
Hambly N. {et~al.}, 2001, MNRAS, 326, 1279

\bibitem[{{Healey} {et~al}\mbox{.}(2008){Healey}, {Romani}, {Cotter},
  {Michelson}, {Schlafly}, {Readhead}, {Giommi}, {Chaty}, {Grenier}, \&
  {Weintraub}}]{Healey2008}
{Healey} S.~E. {et~al.}, 2008, ApJS, 175, 97

\bibitem[{{Heckman} {et~al}\mbox{.}(1994){Heckman}, {O'Dea}, {Baum}, \&
  {Laurikainen}}]{Heckman1994}
{Heckman} T.~M., {O'Dea} C.~P., {Baum} S.~A., {Laurikainen} E., 1994, ApJ, 428,
  65

\bibitem[{{Hewitt} \& {Burbidge}(1987)}]{Hewitt1986}
{Hewitt} A., {Burbidge} G., 1987, ApJs, 63, 1

\bibitem[{{Hickox} \& {Alexander}(2018)}]{Hickox2018}
{Hickox} R.~C., {Alexander} D.~M., 2018, ARAA, 56, 625

\bibitem[{{Hinton} {et~al}\mbox{.}(2016){Hinton}, {Davis}, {Lidman},
  {Glazebrook}, \& {Lewis}}]{Hinton2016}
{Hinton} S.~R., {Davis} T.~M., {Lidman} C., {Glazebrook} K., {Lewis} G.~F.,
  2016, Astronomy and Computing, 15, 61

\bibitem[{Holt {et~al}\mbox{.}(2008)Holt, Tadhunter, \& Morganti}]{Holt2008}
Holt J., Tadhunter C.~N., Morganti R., 2008, MNRAS, 387, 639

\bibitem[{{Hook} {et~al}\mbox{.}(2003){Hook}, {Allington-Smith}, {Beard},
  {Crampton}, {Davies}, {Dickson}, {Ebbers}, {Fletcher}, {Jorgensen}, {Jean},
  {Juneau}, {Murowinski}, {Nolan}, {Laidlaw}, {Leckie}, {Marshall}, {Purkins},
  {Richardson}, {Roberts}, {Simons}, {Smith}, {Stilburn}, {Szeto}, {Tierney},
  {Wolff}, \& {Wooff}}]{Hook2003}
{Hook} I. {et~al.}, 2003, in SPIE, Vol. 4841, Instrument Design and Performance
  for Optical/Infrared Ground-based Telescopes, {Iye} M., {Moorwood} A.~F.~M.,
  eds., pp. 1645--1656

\bibitem[{{Hopkins} {et~al}\mbox{.}(2006){Hopkins}, {Hernquist}, {Cox}, {Di
  Matteo}, {Robertson}, \& {Springel}}]{Hopkins2006}
{Hopkins} P.~F., {Hernquist} L., {Cox} T.~J., {Di Matteo} T., {Robertson} B.,
  {Springel} V., 2006, ApJSS, 163, 1

\bibitem[{{Hopkins} {et~al}\mbox{.}(2008){Hopkins}, {Hernquist}, {Cox}, \&
  {Kere{\v s}}}]{Hopkins2008}
{Hopkins} P.~F., {Hernquist} L., {Cox} T.~J., {Kere{\v s}} D., 2008, ApJS, 175,
  356

\bibitem[{{Hotan} {et~al}\mbox{.}(2014){Hotan}, {Bunton}, {Harvey-Smith},
  {Humphreys}, {Jeffs}, {Shimwell}, {Tuthill}, {Voronkov}, {Allen}, {Amy},
  {Ardern}, {Axtens}, {Ball}, {Bannister}, {Barker}, {Bateman}, {Beresford},
  {Bock}, {Bolton}, {Bowen}, {Boyle}, {Braun}, {Broadhurst}, {Brodrick},
  {Brooks}, {Brothers}, {Brown}, {Cantrall}, {Carrad}, {Chapman}, {Cheng},
  {Chippendale}, {Chung}, {Cooray}, {Cornwell}, {Davis}, {de Souza}, {DeBoer},
  {Diamond}, {Edwards}, {Ekers}, {Feain}, {Ferris}, {Forsyth}, {Gough},
  {Grancea}, {Gupta}, {Guzman}, {Hampson}, {Haskins}, {Hay}, {Hayman}, {Hoyle},
  {Jacka}, {Jackson}, {Jackson}, {Jeganathan}, {Johnston}, {Joseph}, {Kendall},
  {Kesteven}, {Kiraly}, {Koribalski}, {Leach}, {Lenc}, {Lensson}, {Li},
  {Mackay}, {Macleod}, {Maher}, {Marquarding}, {McClure-Griffiths},
  {McConnell}, {Mickle}, {Mirtschin}, {Norris}, {Neuhold}, {Ng}, {O'Sullivan},
  {Pathikulangara}, {Pearce}, {Phillips}, {Qiao}, {Reynolds}, {Rispler},
  {Roberts}, {Roxby}, {Schinckel}, {Shaw}, {Shields}, {Storey}, {Sweetnam},
  {Troup}, {Turner}, {Tzioumis}, {Westmeier}, {Whiting}, {Wilson}, {Wilson},
  {Wormnes}, \& {Wu}}]{Hotan2014}
{Hotan} A.~W. {et~al.}, 2014, PASA, 31, e041

\bibitem[{{Hurley-Walker} {et~al}\mbox{.}(2017){Hurley-Walker}, {Callingham},
  {Hancock}, {Franzen}, {Hindson}, {Kapi{\'n}ska}, {Morgan}, {Offringa},
  {Wayth}, {Wu}, {Zheng}, {Murphy}, {Bell}, {Dwarakanath}, {For}, {Gaensler},
  {Johnston-Hollitt}, {Lenc}, {Procopio}, {Staveley-Smith}, {Ekers}, {Bowman},
  {Briggs}, {Cappallo}, {Deshpande}, {Greenhill}, {Hazelton}, {Kaplan},
  {Lonsdale}, {McWhirter}, {Mitchell}, {Morales}, {Morgan}, {Oberoi}, {Ord},
  {Prabu}, {Shankar}, {Srivani}, {Subrahmanyan}, {Tingay}, {Webster},
  {Williams}, \& {Williams}}]{Hurley-Walker2017}
{Hurley-Walker} N. {et~al.}, 2017, MNRAS, 464, 1146

\bibitem[{{Ishwara-Chandra} {et~al}\mbox{.}(2003){Ishwara-Chandra},
  {Dwarakanath}, \& {Anantharamaiah}}]{Ishwara-Chandra2003}
{Ishwara-Chandra} C.~H., {Dwarakanath} K.~S., {Anantharamaiah} K.~R., 2003,
  Journal of Astrophysics and Astronomy, 24, 37

\bibitem[{{Jackson} {et~al}\mbox{.}(2002){Jackson}, {Wall}, {Shaver},
  {Kellermann}, {Hook}, \& {Hawkins}}]{Jackson2002}
{Jackson} C.~A., {Wall} J.~V., {Shaver} P.~A., {Kellermann} K.~I., {Hook}
  I.~M., {Hawkins} M.~R.~S., 2002, A\&A, 386, 97

\bibitem[{Johnston {et~al}\mbox{.}(2009)Johnston, Feain, \&
  Gupta}]{Johnston2009}
Johnston S., Feain I.~J., Gupta N., 2009, The Low-Frequency Radio Universe ASP
  Conference Series, 407, 446

\bibitem[{Jones {et~al}\mbox{.}(2009)Jones, Read, Saunders, Colless, Jarrett,
  Parker, Fairall, Mauch, Sadler, Watson, Burton, Campbell, Cass, Croom, Dawe,
  Fiegert, Frankcombe, Hartley, Huchra, James, Kirby, Lahav, Lucey, Mamon,
  Moore, Peterson, Prior, Proust, Russell, Safouris, Wakamatsu, Westra, \&
  Williams}]{Jones2009}
Jones D.~H. {et~al.}, 2009, MNRAS, 399, 683

\bibitem[{{Kanekar} \& {Briggs}(2003)}]{Kanekar2003}
{Kanekar} N., {Briggs} F.~H., 2003, A\&A, 412, L29

\bibitem[{{Kanekar} {et~al}\mbox{.}(2014){Kanekar}, {Prochaska}, {Smette},
  {Ellison}, {Ryan-Weber}, {Momjian}, {Briggs}, {Lane}, {Chengalur},
  {Delafosse}, {Grave}, {Jacobsen}, \& {de Bruyn}}]{Kanekar2014}
{Kanekar} N. {et~al.}, 2014, MNRAS, 438, 2131

\bibitem[{{Kollgaard} {et~al}\mbox{.}(1995){Kollgaard}, {Feigelson},
  {Laurent-Muehleisen}, {Spinrad}, {Dey}, \& {Brinkmann}}]{Kollgaard1995}
{Kollgaard} R.~I., {Feigelson} E.~D., {Laurent-Muehleisen} S.~A., {Spinrad} H.,
  {Dey} A., {Brinkmann} W., 1995, ApJ, 449, 61

\bibitem[{{Koss} {et~al}\mbox{.}(2010){Koss}, {Mushotzky}, {Veilleux}, \&
  {Winter}}]{Koss2010}
{Koss} M., {Mushotzky} R., {Veilleux} S., {Winter} L., 2010, ApJL, 716, L125

\bibitem[{K\"{u}hr {et~al}\mbox{.}(1981)K\"{u}hr, Witzel, Pauliny-Toth, \&
  Nauber}]{Kuhr1981}
K\"{u}hr H., Witzel A., Pauliny-Toth I., Nauber U., 1981, A\&ASS, 367

\bibitem[{{Kulkarni} {et~al}\mbox{.}(1997){Kulkarni}, {Fall}, \&
  {Truran}}]{Kulkarni1997}
{Kulkarni} V.~P., {Fall} S.~M., {Truran} J.~W., 1997, ApJL, 484, L7

\bibitem[{{Labiano} {et~al}\mbox{.}(2007){Labiano}, {Barthel}, {O'Dea}, {de
  Vries}, {P{\'e}rez}, \& {Baum}}]{Labiano2007}
{Labiano} A., {Barthel} P.~D., {O'Dea} C.~P., {de Vries} W.~H., {P{\'e}rez} I.,
  {Baum} S.~A., 2007, AAP, 463, 97

\bibitem[{{Lang} {et~al}\mbox{.}(2010){Lang}, {Hogg}, {Mierle}, {Blanton}, \&
  {Roweis}}]{Lang2010}
{Lang} D., {Hogg} D.~W., {Mierle} K., {Blanton} M., {Roweis} S., 2010, AJ, 139,
  1782

\bibitem[{{Ledoux} {et~al}\mbox{.}(2015){Ledoux}, {Noterdaeme}, {Petitjean}, \&
  {Srianand}}]{Ledoux2015}
{Ledoux} C., {Noterdaeme} P., {Petitjean} P., {Srianand} R., 2015, A\&AP, 580,
  A8

\bibitem[{{Lee} {et~al}\mbox{.}(2013){Lee}, {Hwang}, \& {Ko}}]{Lee2013}
{Lee} J.~C., {Hwang} H.~S., {Ko} J., 2013, ApJ, 774, 62

\bibitem[{{Lonsdale} {et~al}\mbox{.}(2016){Lonsdale}, {Whittle}, {Trapp},
  {Patil}, {Lonsdale}, {Thorp}, {Lacy}, {Kimball}, {Blain}, {Jones}, \&
  {Kim}}]{Lonsdale2016}
{Lonsdale} C.~J. {et~al.}, 2016, Astronomische Nachrichten, 337, 194

\bibitem[{{Maccagni} {et~al}\mbox{.}(2017){Maccagni}, {Morganti}, {Oosterloo},
  {Ger{\'e}b}, \& {Maddox}}]{Maccagni2017}
{Maccagni} F.~M., {Morganti} R., {Oosterloo} T.~A., {Ger{\'e}b} K., {Maddox}
  N., 2017, A\&A, 604, A43

\bibitem[{{Malhotra} {et~al}\mbox{.}(1997){Malhotra}, {Rhoads}, \&
  {Turner}}]{Malhotra1997}
{Malhotra} S., {Rhoads} J.~E., {Turner} E.~L., 1997, MNRAS, 288, 138

\bibitem[{Mauch {et~al}\mbox{.}(2003)Mauch, Murphy, Buttery, Curran, Hunstead,
  Piestrzynski, Robertson, \& Sadler}]{Mauch2003}
Mauch T., Murphy T., Buttery H.~J., Curran J., Hunstead R.~W., Piestrzynski B.,
  Robertson J.~G., Sadler E.~M., 2003, MNRAS, 342, 1117

\bibitem[{{McCarthy} {et~al}\mbox{.}(1996){McCarthy}, {Kapahi}, {van Breugel},
  {Persson}, {Athreya}, \& {Subrahmanya}}]{McCarthy1996}
{McCarthy} P.~J., {Kapahi} V.~K., {van Breugel} W., {Persson} S.~E., {Athreya}
  R., {Subrahmanya} C.~R., 1996, ApJS, 107, 19

\bibitem[{{McConnell} {et~al}\mbox{.}(2016){McConnell}, {Allison}, {Bannister},
  {Bell}, {Bignall}, {Chippendale}, {Edwards}, {Harvey-Smith}, {Hegarty},
  {Heywood}, {Hotan}, {Indermuehle}, {Lenc}, {Marvil}, {Popping}, {Raja},
  {Reynolds}, {Sault}, {Serra}, {Voronkov}, {Whiting}, {Amy}, {Axtens}, {Ball},
  {Bateman}, {Bock}, {Bolton}, {Brodrick}, {Brothers}, {Brown}, {Bunton},
  {Cheng}, {Cornwell}, {DeBoer}, {Feain}, {Gough}, {Gupta}, {Guzman},
  {Hampson}, {Hay}, {Hayman}, {Hoyle}, {Humphreys}, {Jacka}, {Jackson},
  {Jackson}, {Jeganathan}, {Joseph}, {Koribalski}, {Leach}, {Lensson},
  {MacLeod}, {Mackay}, {Marquarding}, {McClure-Griffiths}, {Mirtschin},
  {Mitchell}, {Neuhold}, {Ng}, {Norris}, {Pearce}, {Qiao}, {Schinckel},
  {Shields}, {Shimwell}, {Storey}, {Troup}, {Turner}, {Tuthill}, {Tzioumis},
  {Wark}, {Westmeier}, {Wilson}, \& {Wilson}}]{McConnell2016}
{McConnell} D. {et~al.}, 2016, PASA, 33, e042

\bibitem[{Meyer {et~al}\mbox{.}(2004)Meyer, Zwaan, Webster, Staveley-Smith,
  Ryan-Weber, Drinkwater, Barnes, Howlett, Kilborn, Stevens, Waugh, Pierce,
  Bhathal, de~Blok, Disney, Ekers, Freeman, Garcia, Gibson, Harnett, Henning,
  Jerjen, Kesteven, Knezek, Koribalski, Mader, Marquarding, Minchin, O'Brien,
  Oosterloo, Price, Putman, Ryder, Sadler, Stewart, Stootman, \&
  Wright}]{Meyer2004}
Meyer M. {et~al.}, 2004, MNRAS, 350, 1195

\bibitem[{{Morganti} {et~al}\mbox{.}(2013){Morganti}, {Frieswijk}, {Oonk},
  {Oosterloo}, \& {Tadhunter}}]{Morganti2013}
{Morganti} R., {Frieswijk} W., {Oonk} R.~J.~B., {Oosterloo} T., {Tadhunter} C.,
  2013, A\&A, 552, L4

\bibitem[{{Morganti} {et~al}\mbox{.}(2002){Morganti}, {Peck}, {Oosterloo},
  {Capetti}, {Parma}, {de Ruiter}, {Fanti}, \& {van Moorsel}}]{Morganti2002}
{Morganti} R., {Peck} A.~B., {Oosterloo} T.~A., {Capetti} A., {Parma} P., {de
  Ruiter} H., {Fanti} R., {van Moorsel} G., 2002, in Proceedings of the 6th EVN
  Symposium, {Ros} E., {Porcas} R.~W., {Lobanov} A.~P., {Zensus} J.~A., eds.,
  p. 171

\bibitem[{{Moss} {et~al}\mbox{.}(2017){Moss}, {Allison}, {Sadler}, {Urquhart},
  {Soria}, {Callingham}, {Curran}, {Musaeva}, {Mahony}, {Glowacki}, {Farrell},
  {Bannister}, {Chippendale}, {Edwards}, {Harvey-Smith}, {Heywood}, {Hotan},
  {Indermuehle}, {Lenc}, {Marvil}, {McConnell}, {Reynolds}, {Voronkov}, {Wark},
  \& {Whiting}}]{Moss2017}
{Moss} V.~A. {et~al.}, 2017, MNRAS, 471, 2952

\bibitem[{Murphy {et~al}\mbox{.}(2010)Murphy, Sadler, Ekers, Massardi, Hancock,
  Mahony, Ricci, Burke-Spolaor, Calabretta, \& Chhetri}]{Murphy2010}
Murphy T. {et~al.}, 2010, MNRAS, 402, 2403

\bibitem[{{Norris} {et~al}\mbox{.}(2006){Norris}, {Afonso}, {Appleton},
  {Boyle}, {Ciliegi}, {Croom}, {Huynh}, {Jackson}, {Koekemoer}, {Lonsdale},
  {Middelberg}, {Mobasher}, {Oliver}, {Polletta}, {Siana}, {Smail}, \&
  {Voronkov}}]{Norris2006}
{Norris} R.~P. {et~al.}, 2006, AJ, 132, 2409

\bibitem[{{Noterdaeme} {et~al}\mbox{.}(2010){Noterdaeme}, {Petitjean},
  {Ledoux}, {L{\'o}pez}, {Srianand}, \& {Vergani}}]{Noterdaeme2010}
{Noterdaeme} P., {Petitjean} P., {Ledoux} C., {L{\'o}pez} S., {Srianand} R.,
  {Vergani} S.~D., 2010, A\&A, 523, A80

\bibitem[{{Noterdaeme} {et~al}\mbox{.}(2009){Noterdaeme}, {Petitjean},
  {Ledoux}, \& {Srianand}}]{Noterdaeme2009}
{Noterdaeme} P., {Petitjean} P., {Ledoux} C., {Srianand} R., 2009, A\&A, 505,
  1087

\bibitem[{{O'Dea} {et~al}\mbox{.}(1991){O'Dea}, {Baum}, \&
  {Stanghellini}}]{ODea1991}
{O'Dea} C.~P., {Baum} S.~A., {Stanghellini} C., 1991, ApJ, 380, 66

\bibitem[{Oosterloo {et~al}\mbox{.}(2009)Oosterloo, Verheijen, van Cappellen,
  Bakker, Heald, \& Ivashina}]{Oosterloo2009}
Oosterloo T., Verheijen M. A.~W., van Cappellen W., Bakker L., Heald G.,
  Ivashina M., 2009, in "Proceedings of Wide Field Astronomy \& Technology for
  the Square Kilometre Array (SKADS 2009). 4-6 November 2009. Chateau de
  Limelette, p.~70

\bibitem[{{Pontzen} \& {Pettini}(2009)}]{Pontzen2009}
{Pontzen} A., {Pettini} M., 2009, MNRAS, 393, 557

\bibitem[{{Rao} {et~al}\mbox{.}(2017){Rao}, {Turnshek}, {Sardane}, \&
  {Monier}}]{Rao2017}
{Rao} S.~M., {Turnshek} D.~A., {Sardane} G.~M., {Monier} E.~M., 2017, MNRAS,
  471, 3428

\bibitem[{Richards {et~al}\mbox{.}(2015)Richards, Myers, Peters, Krawczyk,
  Chase, Ross, Fan, Jiang, Lacy, McGreer, Trump, \& Riegel}]{Richards2015}
Richards G.~T. {et~al.}, 2015, AJ, 219, 39

\bibitem[{{Robitaille} \& {Bressert}(2012)}]{Robitaille2012}
{Robitaille} T., {Bressert} E., 2012, {APLpy: Astronomical Plotting Library in
  Python}. Astrophysics Source Code Library

\bibitem[{{Satyapal} {et~al}\mbox{.}(2014){Satyapal}, {Ellison}, {McAlpine},
  {Hickox}, {Patton}, \& {Mendel}}]{Satyapal2014}
{Satyapal} S., {Ellison} S.~L., {McAlpine} W., {Hickox} R.~C., {Patton} D.~R.,
  {Mendel} J.~T., 2014, MNRAS, 441, 1297

\bibitem[{Schinckel {et~al}\mbox{.}(2012)Schinckel, Bunton, Cornwell, Feain, \&
  Hay}]{Schinckel2012}
Schinckel A.~E., Bunton J.~D., Cornwell T.~J., Feain I., Hay S.~G., 2012, Proc.
  SPIE 8444, Ground-based and Airborne Telescopes IV, 8444, 84442A

\bibitem[{{Snellen} {et~al}\mbox{.}(2002){Snellen}, {Lehnert}, {Bremer}, \&
  {Schilizzi}}]{Snellen2002}
{Snellen} I.~A.~G., {Lehnert} M.~D., {Bremer} M.~N., {Schilizzi} R.~T., 2002,
  MNRAS, 337, 981

\bibitem[{{Srianand} {et~al}\mbox{.}(2015){Srianand}, {Gupta}, {Momjian}, \&
  {Vivek}}]{Srianand2015}
{Srianand} R., {Gupta} N., {Momjian} E., {Vivek} M., 2015, MNRAS, 451, 917

\bibitem[{{Srianand} {et~al}\mbox{.}(2008){Srianand}, {Gupta}, {Petitjean},
  {Noterdaeme}, \& {Saikia}}]{Srianand2008}
{Srianand} R., {Gupta} N., {Petitjean} P., {Noterdaeme} P., {Saikia} D.~J.,
  2008, MNRAS, 391, L69

\bibitem[{{Stanghellini} {et~al}\mbox{.}(1997){Stanghellini}, {O'Dea}, {Baum},
  {Dallacasa}, {Fanti}, \& {Fanti}}]{Stanghellini1997}
{Stanghellini} C., {O'Dea} C.~P., {Baum} S.~A., {Dallacasa} D., {Fanti} R.,
  {Fanti} C., 1997, A\&A, 325, 943

\bibitem[{{Stern} {et~al}\mbox{.}(2012){Stern}, {Assef}, {Benford}, {Blain},
  {Cutri}, {Dey}, {Eisenhardt}, {Griffith}, {Jarrett}, {Lake}, {Masci},
  {Petty}, {Stanford}, {Tsai}, {Wright}, {Yan}, {Harrison}, \&
  {Madsen}}]{Stern2012}
{Stern} D. {et~al.}, 2012, ApJ, 753, 30

\bibitem[{Stickel {et~al}\mbox{.}(1996)Stickel, Rieke, K\"{u}hr, \&
  Rieke}]{Stickel1996}
Stickel M., Rieke G., K\"{u}hr H., Rieke M., 1996, ApJ, 556

\bibitem[{{Storrie-Lombardi} \& {Wolfe}(2000)}]{Storrie-Lombardi2000}
{Storrie-Lombardi} L.~J., {Wolfe} A.~M., 2000, ApJ, 543, 552

\bibitem[{{Tadhunter} {et~al}\mbox{.}(1993){Tadhunter}, {Morganti}, {di
  Serego-Alighieri}, {Fosbury}, \& {Danziger}}]{Tadhunter1993}
{Tadhunter} C.~N., {Morganti} R., {di Serego-Alighieri} S., {Fosbury} R.~A.~E.,
  {Danziger} I.~J., 1993, MNRAS, 263, 999

\bibitem[{van Gorkom {et~al}\mbox{.}(1989)van Gorkom, Knapp, Ekers, Ekers,
  Laing, \& Polk}]{VanGorkom1989}
van Gorkom J.~H., Knapp G.~R., Ekers R.~D., Ekers D.~D., Laing R.~A., Polk
  K.~S., 1989, AJ, 97, 708

\bibitem[{{Vazdekis} {et~al}\mbox{.}(2010){Vazdekis},
  {S{\'a}nchez-Bl{\'a}zquez}, {Falc{\'o}n-Barroso}, {Cenarro}, {Beasley},
  {Cardiel}, {Gorgas}, \& {Peletier}}]{Vazdekis2010}
{Vazdekis} A., {S{\'a}nchez-Bl{\'a}zquez} P., {Falc{\'o}n-Barroso} J.,
  {Cenarro} A.~J., {Beasley} M.~A., {Cardiel} N., {Gorgas} J., {Peletier}
  R.~F., 2010, MNRAS, 404, 1639

\bibitem[{Verheijen {et~al}\mbox{.}(2007)Verheijen, van Gorkom, Szomoru,
  Dwarakanath, Poggianti, \& Schiminovich}]{Verheijen2007}
Verheijen M., van Gorkom J.~H., Szomoru a., Dwarakanath K.~S., Poggianti B.~M.,
  Schiminovich D., 2007, ApJ, 668, L9

\bibitem[{Vermeulen {et~al}\mbox{.}(2003)Vermeulen, Pihlstr\"{o}m, Tschager,
  de~Vries, Conway, Barthel, Baum, Braun, Bremer, \& Miley}]{Vermeulen2003}
Vermeulen R.~C. {et~al.}, 2003, A\&A, 404, 861

\bibitem[{{Webster} {et~al}\mbox{.}(1995){Webster}, {Francis}, {Petersont},
  {Drinkwater}, \& {Masci}}]{Webster1995}
{Webster} R.~L., {Francis} P.~J., {Petersont} B.~A., {Drinkwater} M.~J.,
  {Masci} F.~J., 1995, Nature, 375, 469

\bibitem[{{Weston} {et~al}\mbox{.}(2017){Weston}, {McIntosh}, {Brodwin},
  {Mann}, {Cooper}, {McConnell}, \& {Nielsen}}]{Weston2017}
{Weston} M.~E., {McIntosh} D.~H., {Brodwin} M., {Mann} J., {Cooper} A.,
  {McConnell} A., {Nielsen} J.~L., 2017, MNRAS, 464, 3882

\bibitem[{{White} {et~al}\mbox{.}(1997){White}, {Becker}, {Helfand}, \&
  {Gregg}}]{White1997}
{White} R.~L., {Becker} R.~H., {Helfand} D.~J., {Gregg} M.~D., 1997, ApJ, 475,
  479

\bibitem[{{White} {et~al}\mbox{.}(2003){White}, {Helfand}, {Becker}, {Gregg},
  {Postman}, {Lauer}, \& {Oegerle}}]{White2003}
{White} R.~L., {Helfand} D.~J., {Becker} R.~H., {Gregg} M.~D., {Postman} M.,
  {Lauer} T.~R., {Oegerle} W., 2003, AJ, 126, 706

\bibitem[{{Whiting} {et~al}\mbox{.}(2001){Whiting}, {Webster}, \&
  {Francis}}]{Whiting2001}
{Whiting} M.~T., {Webster} R.~L., {Francis} P.~J., 2001, MNRAS, 323, 718

\bibitem[{{Wright} {et~al}\mbox{.}(1983){Wright}, {Ables}, \&
  {Allen}}]{Wright1983}
{Wright} A.~E., {Ables} J.~G., {Allen} D.~A., 1983, MNRAS, 205, 793

\bibitem[{Wright {et~al}\mbox{.}(2010)Wright, Eisenhardt, Mainzer, Ressler,
  Cutri, Jarrett, Kirkpatrick, Padgett, McMillan, Skrutskie, Stanford, Cohen,
  Walker, Mather, Leisawitz, III, McLean, Benford, Lonsdale, Blain, Mendez,
  Irace, Duval, Liu, Royer, Heinrichsen, Howard, Shannon, Kendall, Walsh,
  Larsen, Cardon, Schick, Schwalm, Abid, Fabinsky, Naes, \& Tsai}]{Wright2010}
Wright E.~L. {et~al.}, 2010, AJ, 140, 1868

\bibitem[{Wu {et~al}\mbox{.}(2012)Wu, Hao, Jia, Zhang, \& Peng}]{Wu2012}
Wu X.-B., Hao G., Jia Z., Zhang Y., Peng N., 2012, AJ, 144, 49

\bibitem[{{Yuan} {et~al}\mbox{.}(2013){Yuan}, {Liu}, \& {Xiang}}]{Yuan2013}
{Yuan} H.~B., {Liu} X.~W., {Xiang} M.~S., 2013, MNRAS, 430, 2188

\end{thebibliography}
}
\label{lastpage}
\end{document}